\documentclass[12pt,english]{article}
\usepackage{times}
\usepackage[T1]{fontenc}
\usepackage{geometry}
\usepackage{color}
\usepackage{graphicx}
\usepackage{setspace}
\usepackage{amssymb}

\makeatletter

\providecommand{\tabularnewline}{\\}

\usepackage{bm}

\usepackage{babel}
\makeatother
\begin{document}

\section*{\noindent Array Antenna Power Pattern Analysis Through Quantum Computing}

\noindent \vfill

\noindent L. Tosi,$^{(1)}$ P. Rocca, $^{(1)(2)}$ \emph{Senior Member,
IEEE}, \emph{}N. Anselmi,$^{(1)}$ \emph{Member, IEEE}, and A. Massa,$^{(1)(3)(4)}$
\emph{Fellow, IEEE}

\noindent \vfill

\noindent {\footnotesize ~}{\footnotesize \par}

\noindent {\footnotesize $^{(1)}$} \emph{\footnotesize ELEDIA Research
Center} {\footnotesize (}\emph{\footnotesize ELEDIA@UNITN} {\footnotesize -
University of Trento)}{\footnotesize \par}

\noindent {\footnotesize Via Mesiano 77, 38123 Trento - Italy}{\footnotesize \par}

\noindent \textit{\emph{\footnotesize E-mail:}} {\footnotesize \{}\emph{\footnotesize luca.tosi-1,
paolo.rocca, nicola.anselmi.1, andrea.massa}{\footnotesize \}@}\emph{\footnotesize unitn.it}{\footnotesize \par}

\noindent {\footnotesize Website:} \emph{\footnotesize www.eledia.org/eledia-unitn}{\footnotesize \par}

\noindent {\footnotesize ~}{\footnotesize \par}

\noindent {\footnotesize $^{(2)}$} \emph{\footnotesize ELEDIA Research
Center} {\footnotesize (}\emph{\footnotesize ELEDIA}{\footnotesize @}\emph{\footnotesize XIDIAN}
{\footnotesize - Xidian University)}{\footnotesize \par}

\noindent {\footnotesize P.O. Box 191, No.2 South Tabai Road, 710071
Xi'an, Shaanxi Province - China }{\footnotesize \par}

\noindent {\footnotesize E-mail:} \emph{\footnotesize paolo.rocca@xidian.edu.cn}{\footnotesize \par}

\noindent {\footnotesize Website:} \emph{\footnotesize www.eledia.org/eledia-xidian}{\footnotesize \par}

\noindent {\footnotesize ~}{\footnotesize \par}

\noindent {\footnotesize $^{(3)}$} \emph{\footnotesize ELEDIA Research
Center} {\footnotesize (}\emph{\footnotesize ELEDIA}{\footnotesize @}\emph{\footnotesize UESTC}
{\footnotesize - UESTC)}{\footnotesize \par}

\noindent {\footnotesize School of Electronic Engineering, Chengdu
611731 - China}{\footnotesize \par}

\noindent \textit{\emph{\footnotesize E-mail:}} \emph{\footnotesize andrea.massa@uestc.edu.cn}{\footnotesize \par}

\noindent {\footnotesize Website:} \emph{\footnotesize www.eledia.org/eledia}{\footnotesize -}\emph{\footnotesize uestc}{\footnotesize \par}

\noindent {\footnotesize ~}{\footnotesize \par}

\noindent {\footnotesize $^{(4)}$} \emph{\footnotesize ELEDIA Research
Center} {\footnotesize (}\emph{\footnotesize ELEDIA@TSINGHUA} {\footnotesize -
Tsinghua University)}{\footnotesize \par}

\noindent {\footnotesize 30 Shuangqing Rd, 100084 Haidian, Beijing
- China}{\footnotesize \par}

\noindent {\footnotesize E-mail:} \emph{\footnotesize andrea.massa@tsinghua.edu.cn}{\footnotesize \par}

\noindent {\footnotesize Website:} \emph{\footnotesize www.eledia.org/eledia-tsinghua}{\footnotesize \par}

\noindent \vfill

\textbf{\emph{This work has been submitted to the IEEE for possible
publication. Copyright may be transferred without notice, after which
this version may no longer be accessible.}}

\noindent \vfill

\newpage
\section*{Array Antenna Power Pattern Analysis Through Quantum Computing}

~

~

~

\begin{flushleft}L. Tosi, P. Rocca, N. Anselmi, and A. Massa\end{flushleft}

\vfill

\begin{abstract}
\noindent A method for the analysis of the power pattern of phased
array antennas (\emph{PA}s) based on the quantum Fourier transform
(\emph{QFT}) is proposed. The computation of the power pattern given
the set of complex excitations of the \emph{PA} elements is addressed
within the quantum computing (\emph{QC}) framework by means of a customized
procedure that exploits the quantum mechanics principles and theory.
A representative set of numerical results, yielded with a quantum
computer emulator, is reported and discussed to assess the reliability
of the proposed method by pointing out its features in comparison
with the classical approach based on the discrete Fourier transform
(\emph{DFT}), as well.

\vfill
\end{abstract}
\noindent \textbf{Key words}: Array Antenna; Antenna Analysis; Power
Pattern; Quantum Computing (\emph{QC}); Quantum Fourier Transform
(\emph{QFT}).

\newpage
\section{Introduction}

\noindent \textcolor{black}{Nowadays, phased array antennas (}\textcolor{black}{\emph{PA}}\textcolor{black}{s)
are a widely adopted technology with countless applications \cite{Mailloux 2018}\cite{Haupt 2010}
such as multiple-input multiple-output (}\textcolor{black}{\emph{MIMO}}\textcolor{black}{)
mobile communications (e.g., 5G) \cite{Mao.2019}-\cite{Puglielli.2016},
space communications \cite{Davarian.2007}-\cite{Moon.2019}, automotive
\cite{Yu.2008} and airborne radar systems \cite{Gottardi.2017},
and weather forecasting \cite{Li.2021} as well as biomedical microwave
imaging \cite{Bucci.2016}\cite{Tofigh.2014} or, more in general,
non destructive evaluation and testing (}\textcolor{black}{\emph{NDE}}\textcolor{black}{/}\textcolor{black}{\emph{NDT}}\textcolor{black}{)
\cite{Li.2019}. Indeed,} \textcolor{black}{\emph{PA}}\textcolor{black}{s
stand out for the robustness, the adaptability to conformal surfaces,
the easy reconfiguration capabilities, and the careful control of
the radiation features \cite{Herd.2016}. Moreover, it is worth mentioning
that unconventional (i.e., clustered \cite{Benoni.2022}\cite{Rocca.2021.b}
or thinned/sparse \cite{Oliveri.2009}\cite{Poli.2013} architectures)}
\textcolor{black}{\emph{PA}}\textcolor{black}{s are gaining momentum
since they are interesting solutions as suitable trade-offs between
costs and radiation performance \cite{Rocca.2016}.}

\noindent \textcolor{black}{Typically, the design of} \textcolor{black}{\emph{PA}}\textcolor{black}{s
has been addressed by formulating the synthesis problem at hand as
an optimization one, then solved with deterministic \cite{Rocca.2014}-\cite{Yilmaz.2017}
or stochastic \cite{Lin.2010}\cite{Angeletti.2022} methods by recurring
to machine learning techniques \cite{Cui.2022}, as well. However,
a common bottleneck for many design methods is the need of iterating
the evaluation of a cost function, which is based on the knowledge/computation
of the power pattern, that quantifies the mismatch between synthesized
pattern features and the ideal one or against user-defined requirements.
Therefore, the computational cost becomes prohibitive in case of large
arrays \cite{Rocca.2014.b} because of the wide dimension of the solution
space as well as the huge number of pattern samples to be computed
for properly accounting for its angular variations.}

\noindent \textcolor{black}{The goal of this work is to propose an
innovative method, based on} \textcolor{black}{\emph{QC}}\textcolor{black}{,
for the fast computation of the power pattern starting from the array
excitations (also indicated as {}``}\textcolor{black}{\emph{PAs analysis}}\textcolor{black}{''
problem). Indeed,} \textcolor{black}{\emph{QC}} \textcolor{black}{has
the potential for offering a huge computational power, unreachable
even by current supercomputers.} \textcolor{black}{\emph{}}\textcolor{black}{In
the recent years,} \textcolor{black}{\emph{QC}} \textcolor{black}{has
drawn significant attention since it has proved the ability to solve
computationally-intensive problems by taking advantage of phenomena
exclusively related to the quantum realm. For instance, the superposition
and the entanglement have been used in many quantum algorithms to
enable a relevant speedup versus classical programs \cite{Deutsch.1992}.
On the other hand, a widespread use of quantum computers is still
prevented by different technical factors such as the high errors intrinsic
to the quantum gates \cite{Nielsen 2000}, which are used to perform
quantum operations, and the instability when dealing with large numbers
of quantum bits (qubits) \cite{Wilde 2017}. Despite these technological
issues,} \textcolor{black}{\emph{QC}} \textcolor{black}{has been already
exploited in different research branches starting from physics and
computational biology \cite{McArdle.2020} up to, more recently, computational
electromagnetics \cite{Rocca.2021.a}\cite{Ross.2021} for the simulation
and optimization of complex radiating systems. Undoubtedly, the most
well-known applications of} \textcolor{black}{\emph{QC}} \textcolor{black}{are
in the field of information technology as, for instance, quantum machine
learning \cite{Schuld.2020} and quantum cryptography \cite{Xu.2015}.
In this latter framework, a very popular} \textcolor{black}{\emph{QC}}
\textcolor{black}{algorithm is the Shor's algorithm \cite{Shor 1999}
that has rendered ineffective the Rivest-Shamir-Adleman (}\textcolor{black}{\emph{RSA}}\textcolor{black}{)
cryptography technique \cite{Rivest.1978} by solving the underlying
factoring problem in polynomial time. Moreover, a part of the Shor's
algorithm, namely the quantum Fourier transform (}\textcolor{black}{\emph{QFT}}\textcolor{black}{),
has given an exponential advantage in terms of computational efficiency
over its classical counterpart {[}i.e., the discrete Fourier transform
(}\textcolor{black}{\emph{DFT}}\textcolor{black}{){]}.}

\noindent \textcolor{black}{In this paper, for the first time to the
best of the authors' knowledge, the prediction of the power pattern
of} \textcolor{black}{\emph{PA}}\textcolor{black}{s is efficiently
carried out by benefiting of the lower computational costs of the}
\textcolor{black}{\emph{QFT}}\textcolor{black}{. Towards this end,
the} \textcolor{black}{\emph{PA}}\textcolor{black}{s} \textcolor{black}{\emph{analysis
problem}} \textcolor{black}{is adapted to the quantum domain since,
while quantum computers are able to perform all the operations available
on classical computers \cite{Rieffel.2014}, it is mandatory to reformulate
the problem at hand to reach the} \textcolor{black}{\emph{QC}} \textcolor{black}{speedup.
More in detail, it means firstly to encode the classical information
into the quantum one for taking advantage of the set of} \textcolor{black}{\emph{QC}}
\textcolor{black}{exclusive operations. Moreover, a user of a quantum
computer needs to know that the meaning of output data radically differs
from the classical one. Indeed, each execution of a quantum program
results in the measurement of a quantum state and it is possible to
derive the probability of such a state, which is here strictly related
to the result of the Fourier transform operation, only by iterating
the execution of the program for a number of runs (indicated as '}\textcolor{black}{\emph{shots}}\textcolor{black}{').}

\noindent \textcolor{black}{The rest of the paper is organized as
follows. In Sect. 2, the} \textcolor{black}{\emph{PA}} \textcolor{black}{analysis
problem is mathematically formulated in the} \textcolor{black}{\emph{QC}}
\textcolor{black}{framework by presenting the} \textcolor{black}{\emph{QFT}}\textcolor{black}{-based
pattern prediction method. Section 3 is devoted to the numerical validation
and assessment of the proposed computational approach. Representative
numerical results are also provided to give some insights on the behavior
of the} \textcolor{black}{\emph{QFT}} \textcolor{black}{algorithm
and the dependence of its performance on the control parameters. Eventually,
some conclusions and final remarks are drawn (Sect. 4).}

\section{\noindent \textcolor{black}{Mathematical Formulation\label{sec:Mathematical-Formulation}}}

\noindent \textcolor{black}{Let us consider a $N$ elements} \textcolor{black}{\emph{PA}}
\textcolor{black}{where the array elements are equally-spaced by $d$
along the $z$-axis (Fig. 1). Each $n$-th ($n=0,...,N-1$) element
is connected to a transmit-receive module (}\textcolor{black}{\emph{TRM}}\textcolor{black}{)
that generates a complex excitation $w_{n}$, $w_{n}\triangleq\alpha_{n}\textrm{exp}\left(j\beta_{n}\right)$,
($n=0,...,N-1$), $\alpha_{n}$ and $\beta_{n}$ being the corresponding
amplitude and phase, respectively, while $j=\sqrt{-1}$ is the imaginary
unit. Let $\underline{g}_{n}\left(u\right)$ be the radiation pattern
of the $n$-th ($n=0,...,N-1$) array element, then the} \textcolor{black}{\emph{EM}}
\textcolor{black}{field radiated in far-field} (\emph{FF}) \textcolor{black}{by
the} \textcolor{black}{\emph{PA}} \textcolor{black}{is given by\begin{equation}
\underline{F}\left(u\right)=\sum_{n=0}^{N-1}w_{n}\underline{g}_{n}\left(u\right)\textrm{exp}\left[j\left(kdu\right)n\right]\label{eq:_radiated.field}\end{equation}
where $k=2\pi/\lambda$ is the free-space wavenumber, $\lambda$ being
the corresponding wavelength, and $u$ is the direction cosine ($u\triangleq\cos\theta$),
$\theta$ ($0\le\theta\le\pi$) being the angle measured from the
$z$-axis. If the array elements are equal {[}i.e., $\underline{g}_{n}\left(u\right)\simeq\underline{g}\left(u\right)$;
$n=0,...,N-1${]}, the $\underline{F}\left(u\right)$ can be written
as the product between the element pattern, $\underline{g}\left(u\right)$,
and the array factor, $A\left(u\right)$\begin{equation}
A\left(u\right)\triangleq\sum_{n=0}^{N-1}w_{n}\textrm{exp}\left[j\left(kdu\right)n\right],\label{eq:_array.factor}\end{equation}
{[}i.e., $\underline{F}(u)=\underline{g}\left(u\right)A\left(u\right)${]},
and the corresponding power pattern, $P\left(u\right)$ ($P\left(u\right)\triangleq\left|\underline{F}\left(u\right)\right|^{2}$),
which mathematically describes the angular distribution of the power
either radiated or received by the} \textcolor{black}{\emph{PA}}\textcolor{black}{,
turns out to be\begin{equation}
P\left(u\right)=\left|\underline{g}\left(u\right)\right|^{2}\left|A\left(u\right)\right|^{2}.\label{eq:_power.pattern}\end{equation}
As it can be observed (\ref{eq:_power.pattern}), $P\left(u\right)$
directly depends on the absolute square of the array factor, $\left|A\left(u\right)\right|^{2}$,
and the samples of this latter (\ref{eq:_array.factor}), $\mathbf{A}=\left\{ A_{m};\, m=0,\dots,M-1\right\} $,
are related to the set of the array excitations, $\mathbf{w}=\left\{ w_{n};\, n=0,\dots,N-1\right\} $,
through the} \textcolor{black}{\emph{DFT}}\textcolor{black}{, $\mathbf{A}=DFT\left(\mathbf{w}\right)$\begin{equation}
A_{m}=\sum_{n=0}^{N-1}w_{n}\textrm{exp}\left[-j\left(\frac{2\pi}{N}n\right)m\right],\,\,\,\,\, m=0,\dots,M-1\label{eq:_AF.DFT}\end{equation}
where $A_{m}=A\left(u_{m}\right)$ is the $m$-th ($m=0,\dots,M-1$)
sample of the array factor (\ref{eq:_array.factor}) at the angular
direction $u_{m}=-m\frac{\lambda}{Nd}$.}

\noindent \textcolor{black}{Once the discrete sample vector $\mathbf{A}$
has been computed, the corresponding continuous function, \{$A\left(u\right)$;
$-1\le u\le1$\}, is obtained by means of a periodic interpolation\begin{equation}
A\left(u\right)=\sum_{m=0}^{M-1}A_{m}S\left(\pi du+\frac{m\pi}{N}\right),\label{eq:_AF.interpolation}\end{equation}
$S$ being the sinc function {[}$S\left(x\right)\triangleq\textrm{sin}\left(Nx\right)/N\textrm{sin}\left(x\right)${]}.}

\noindent \textcolor{black}{Under the hypothesis of ideal elements
{[}i.e., $\underline{g}\left(u\right)=\underline{1}${]} through (\ref{eq:_power.pattern}),
it turns out that $P\left(u\right)=\left|A\left(u\right)\right|^{2}$,
thus the relation \begin{equation}
P_{m}=\left|A_{m}\right|^{2}\label{eq:_power.pattern.DFT}\end{equation}
 holds true, $P_{m}$ being the sample of the power patter at $u=u_{m}$
{[}i.e., $P_{m}=P\left(u_{m}\right)${]}.}

\noindent \textcolor{black}{The power pattern} \textcolor{black}{\emph{}}\textcolor{black}{of
the $N$ elements array antenna is then computed in the} \textcolor{black}{\emph{QC}}
\textcolor{black}{framework with the quantum counterpart of the} \textcolor{black}{\emph{DFT}}\textcolor{black}{,
namely the} \textcolor{black}{\emph{QFT}}\textcolor{black}{. Towards
this end, the first step consists in allocating a register of qubits,
$\bm{\Psi}$, where the input/output values of the Fourier transform
are encoded in a set of $Q$ quantum states. More specifically, $\bm{\Psi}$
is the concatenation of $L$ single qubits, each $l$-th ($l=0,...,L-1$)
one assuming two possible states, $\left|\psi_{l}\right\rangle \in\left\{ \left|0\right\rangle ,\left|1\right\rangle \right\} $,
and the $q$-th ($q=0,...,Q-1$) state of the vector $\bm{\Psi}$,
$\bm{\Psi}\leftarrow\left|\bm{\Psi}_{q}\right\rangle $, is given
by the tensor product of each single qubit state $\left|\bm{\Psi}_{q}\right\rangle =\bigotimes_{l=0}^{L-1}\left|\psi_{l}^{q}\right\rangle =\left|\psi_{L-1}^{q}\,...\,\psi_{l}^{q}\,...\,\psi_{0}^{q}\right\rangle $.
Since a $L$ qubit register, $\bm{\Psi}$, can encode at most $2^{L}$
different (input/output) states, then a register with $L=\left\lceil \textrm{log}_{2}Q\right\rceil $,
$Q$ being $Q=\max\left(N,\, M\right)$, qubits must be chosen to
yield, through} \textcolor{black}{\emph{QFT}}\textcolor{black}{, $M$
samples of the power pattern starting from $N$ array excitations.}

\noindent \textcolor{black}{The input state vector of the} \textcolor{black}{\emph{QFT}}
\textcolor{black}{is then initialized by assigning a $n$-th ($n=0,...,N-1$)
excitation, $w_{n}$, to each $q$-th ($q=0,...,Q-1$) quantum state,
$\left|\bm{\Psi}_{q}\right\rangle $. More in detail, the first $N$
states of $\bm{\Psi}$ are assigned to the $N$-size set of normalized
complex excitations $\mathbf{\hat{w}}$, $\mathbf{\hat{w}}\triangleq\left\{ \hat{w}_{n}\triangleq\frac{w_{n}}{\left\Vert \mathbf{w}\right\Vert };\, n=0,...,N-1\right\} $,}%
\footnote{\textcolor{black}{The normalization implies that the sum of the squared
modulus of all the weights is unitary (i.e., $\sum_{n=0}^{N-1}\left|\hat{w}_{n}\right|^{2}=1$).}%
} \textcolor{black}{while the remaining $Q-N$ ones are set to zero}

\noindent \textcolor{black}{\begin{equation}
\left|\mathbf{w}\right\rangle =\sum_{q=0}^{N-1}\hat{w}_{q}\left|\bm{\Psi}_{q}\right\rangle +\sum_{q=N}^{Q-1}0\left|\bm{\Psi}_{q}\right\rangle .\label{eq:_weight.initialization}\end{equation}
Analogously to the classical theory, the output state vector $\left|\mathbf{A}\right\rangle $\begin{equation}
\left|\mathbf{A}\right\rangle =\sum_{m=0}^{M-1}\hat{A}_{m}\left|\bm{\Psi}_{m}\right\rangle \label{eq:_QFT.2}\end{equation}
is yielded by applying the} \textcolor{black}{\emph{QFT}} \textcolor{black}{to
the input quantum state vector $\left|\mathbf{w}\right\rangle $ so
that the weight of the state $\left|\bm{\Psi}_{m}\right\rangle $
($m=0,...,M-1$) is given by\begin{equation}
\hat{A}_{m}=\frac{1}{\sqrt{M}}\sum_{q=0}^{M-1}\hat{w}_{q}\textrm{exp}\left[-j\left(\frac{2\pi}{N}q\right)m\right]\left|\bm{\Psi}_{q}\right\rangle .\label{eq:_QFT}\end{equation}
Unfortunately, the measurable output of a} \textcolor{black}{\emph{QC}}
\textcolor{black}{operation, as the} \textcolor{black}{\emph{QFT}}
\textcolor{black}{one, is not a variable {[}e.g., here the complex
value of $\hat{A}_{m}$ ($m=0,\dots,M-1$){]}, but the probability
of occurrence of the quantum state associated to it {[}i.e., $\left|\bm{\Psi}_{m}\right\rangle $
($m=0,\dots,M-1$){]}. On the other hand, the Born rule \cite{Rieffel.2014}
states that the probability of measuring the $m$-th ($m=0,\dots,M-1$)
quantum state, $\wp_{m}$ , is related to the corresponding $m$-th
weight, $\hat{A}_{m}$, as follows\begin{equation}
\wp_{m}=\left|\hat{A}_{m}\right|^{2}.\label{eq:_Born.rule}\end{equation}
By comparing (\ref{eq:_power.pattern.DFT}) and (\ref{eq:_Born.rule}),
it is easy to infer that\begin{equation}
\hat{\wp}_{m}=\hat{P}_{m}\label{eq:_normalized.values}\end{equation}
where $\hat{\wp}_{m}\triangleq\frac{\wp_{m}}{\wp_{MAX}}$ and $\hat{P}_{m}\triangleq\frac{P_{m}}{P_{MAX}}$
, $\wp_{MAX}$ and $P_{MAX}$ being the normalization coefficients,
$\wp_{MAX}\triangleq\textrm{max}_{m=0,...,M-1}\left(\wp_{m}\right)$
and $P_{MAX}\triangleq\textrm{max}_{m=0,...,M-1}\left(P_{m}\right)$,
respectively.}

\noindent \textcolor{black}{Accordingly, the $m$-th ($m=0,...,M-1$)
sample of the power pattern of the array, $\hat{A}_{m}$, when feeding
the array elements with the excitation vector $\mathbf{\hat{w}}$
is directly obtained by setting its value to the probability of occurrence
of each $m$-th ($m=0,...,M-1$) output vector state, $\left|\bm{\Psi}_{m}\right\rangle $
(\ref{eq:_normalized.values}) without other maths (e.g., the squared
modulus) as in the classical} \textcolor{black}{\emph{DFT}} \textcolor{black}{approach
(\ref{eq:_power.pattern.DFT}).}

\noindent \textcolor{black}{On the other hand, since the output of
a quantum computer is the index of a quantum state, the probability
of measuring the $m$-th ($m=0,...,M-1$) output quantum state, $\left|\bm{\Psi}_{m}\right\rangle $,
is determined by repeatedly ($T$ being the number of executions or
'}\textcolor{black}{\emph{shots}}\textcolor{black}{') running the
same program as\begin{equation}
\wp_{m}=\frac{V_{m}}{T}\label{eq:_probability}\end{equation}
where $V_{m}$ ($m=0,...,M-1$) is the number of times the $m$-th
state $\left|\bm{\Psi}_{m}\right\rangle $ has been observed/measured.
In a noiseless environment, the value of $\wp_{m}$ (\ref{eq:_probability})
converges to the actual probability when $T\rightarrow\infty$. However,
$T$ is a finite number in real cases and it is chosen as a compromise
between the computation time and the accuracy of the arising power
pattern samples.}

\section{\noindent \textcolor{black}{Numerical Assessment}}

\noindent \textcolor{black}{In this section, the proposed} \textcolor{black}{\emph{QFT}}\textcolor{black}{-based
method for the computation of the power patterns of antenna arrays
is assessed by considering various arrays affording different} \textcolor{black}{\emph{FF}}
\textcolor{black}{beam-patterns. For comparison purposes, the power
patterns computed with the classical} \textcolor{black}{\emph{DFT}}\textcolor{black}{-based
approach have been set as reference. The} \textcolor{black}{\emph{QC}}
\textcolor{black}{computations have been emulated with the open Python
library Qiskit \cite{Aleksandrowicz.2019} from IBM \cite{IBM} that
faithfully mimics the behavior of a real quantum computer. In all
simulations, ideal isotropic antennas have been assumed {[}i.e., $\underline{g}_{n}\left(u\right)=\underline{g}\left(u\right)=\underline{1}$
($n=0,...,N-1$){]} to avoid any bias related to the type of the radiating
elements of the array.}

\noindent \textcolor{black}{The first example deals with a linear
array of $N=16$ elements, spaced by $d=\frac{\lambda}{2}$, having
real-valued excitations drawn from a Dolph-Chebychev (}\textcolor{black}{\emph{DC}}\textcolor{black}{)
distribution and affording a power pattern with sidelobe level (}\textcolor{black}{\emph{SLL}}\textcolor{black}{)
equal to $SLL=-15$ {[}dB{]}. The power pattern has been computed
in $M=1024$ angular samples, thus $L=10$ qubits have been allocated
for the} \textcolor{black}{\emph{QC}} \textcolor{black}{process. According
to the guidelines in Sect. \ref{sec:Mathematical-Formulation}, the
complex {}``amplitudes'' of the first $N$ quantum states of $\left|\mathbf{w}\right\rangle $
have been initialized with the} \textcolor{black}{\emph{DC}} \textcolor{black}{excitations
in Tab. I, while the others $Q-N=1008$ entries have been set to zero
as indicated by (\ref{eq:_weight.initialization}). The} \textcolor{black}{\emph{QFT}}
\textcolor{black}{algorithm has been then executed $T=M\times10^{3}$
times and the measured quantum state probabilities, $\left\{ \hat{\wp}_{m};\, m=0,...,M-1\right\} $
being $\wp_{MAX}=1.418\times10^{-2}$, are shown in Fig. 2 along with
the reference pattern yielded by interpolating the patter samples
from the classical} \textcolor{black}{\emph{DFT}}\textcolor{black}{.
One can observe that the} \textcolor{black}{\emph{QFT}} \textcolor{black}{samples
almost perfectly fit the} \textcolor{black}{\emph{DFT}} \textcolor{black}{curve
by assessing the reliability of the} \textcolor{black}{\emph{QC}}
\textcolor{black}{pattern-prediction tool.}

\noindent \textcolor{black}{As for the computational complexity of
the proposed analysis method, it is limited to that of the} \textcolor{black}{\emph{QFT}}
\textcolor{black}{and it amounts to $\Delta_{\mathbb{Q}}=\mathcal{O}\left(\textrm{log}^{2}M\right)$,
while the pattern prediction with the classical} \textcolor{black}{\emph{DFT}}
\textcolor{black}{method needs $\mathcal{O}\left(M\times\textrm{log}M\right)$
and $\mathcal{O}\left(M\right)$ operations {[}i.e., $\Delta_{\mathbb{C}}$
$=$ $\mathcal{O}\left(M\times\left(\textrm{log}M+1\right)\right)${]}
for the fast Fourier transform (}\textcolor{black}{\emph{FFT}}\textcolor{black}{)
and the square of the amplitudes, respectively. Thus, the computational
saving thanks to the} \textcolor{black}{\emph{QC}} \textcolor{black}{turns
out to be $\frac{\Delta_{\mathbb{C}}}{\Delta_{\mathbb{Q}}}>\frac{M}{\textrm{log}M}$.}

\noindent \textcolor{black}{Figure 2 also shows that the samples of
the} \textcolor{black}{\emph{QFT}} \textcolor{black}{are always above
a minimum 'resolution' threshold $\delta$ defined as \begin{equation}
\delta=\frac{1}{V_{MAX}}\label{eq:_delta.resolution}\end{equation}
where $V_{MAX}$ is the maximum number of times the most recurring
output quantum state has been measured (i.e., $V_{MAX}\triangleq\arg\left\{ \textrm{max}_{m}\left(V_{m}\right)\right\} $).
In this case, $\delta=-41.6$ {[}dB{]} being $V_{MAX}=1.4518\times10^{2}$. }

\noindent Since $V_{MAX}$ is expected to statistically grow with
$T$, \textcolor{black}{the value of the resolution threshold $\delta$
should get smaller and smaller (i.e., samples with lower amplitudes
can be observed) when more shots are used for the quantum computation.
To assess such a relation between $T$ and $\delta$, the} \textcolor{black}{\emph{QFT}}
\textcolor{black}{pattern prediction has been performed by varying
$T$, while keeping the same} \textcolor{black}{\emph{DC}} \textcolor{black}{input
state vector, $\mathbf{\left|w\right\rangle }$, of the previous example.
In particular, the $M=1024$ pattern samples in Fig. 3 have been computed
using $T=M\times8$ {[}Fig. 3(}\textcolor{black}{\emph{a}}\textcolor{black}{){]},
$T=M\times20$ {[}Fig. 3(}\textcolor{black}{\emph{b}}\textcolor{black}{){]},
$T=M\times40$ {[}Fig. 3(}\textcolor{black}{\emph{c}}\textcolor{black}{){]},
and $T=M\times80$ {[}Fig. 3(}\textcolor{black}{\emph{d}}\textcolor{black}{){]}
shots and the smallest representable values of the power patterns
turn out to be equal to $\delta=-21.5$ {[}dB{]}, $\delta=-25.0$
{[}dB{]}, $\delta=-27.8$ {[}dB{]}, and $\delta=-30.9$ {[}dB{]},
respectively. Moreover, to provide a statistically reliable assessment,
each simulation has been repeated $R=20$ times and Figure 4 shows
the behavior of the average (solid line) value of $\delta$ along
with its minimum and maximum bounds (shaded region). As expected,
the plot confirms the monotonic decreasing dependence of $\delta$
on $T$.}

\noindent \textcolor{black}{In the third numerical experiment, the
dependence of the accuracy of the} \textcolor{black}{\emph{QFT}}\textcolor{black}{-based
analysis method on the} \textcolor{black}{\emph{SLL}} \textcolor{black}{of
the reference pattern has been evaluated still considering the $N=16$
$d=\frac{\lambda}{2}$-spaced array, but with the excitations in Tab.
I affording the three} \textcolor{black}{\emph{DC}} \textcolor{black}{patterns
having $SLL=\left\{ -15,\,-20,\,-25\right\} $ {[}dB{]}. Figure 5
summarizes the outcomes of the} \textcolor{black}{\emph{QFT}} \textcolor{black}{prediction
process ($M=1024$ and $L=10$) when applied to the $SLL=-20$ {[}dB{]}
and the $SLL=-25$ {[}dB{]} cases by setting $T=M\times80$ shots
as in Fig. 3(}\textcolor{black}{\emph{d}}\textcolor{black}{) ($SLL=-15$
{[}dB{]}). From the comparison, it turns out that there is a growing
number of misplaced samples around the sidelobe peaks when lowering
the} \textcolor{black}{\emph{SLL}} \textcolor{black}{from $SLL=-15$
{[}dB{]} {[}Fig. 3(}\textcolor{black}{\emph{d}}\textcolor{black}{){]}
down to $SLL=-25$ {[}dB{]} {[}Fig. 5(}\textcolor{black}{\emph{b}}\textcolor{black}{){]}
since the prediction of the samples with smaller power pattern values
needs more and more shots ($T\,\uparrow$) as pointed out by the analysis
in Fig. 3.}

\noindent \textcolor{black}{To quantify the mismatch between the reference}
\textcolor{black}{\emph{DFT}} \textcolor{black}{pattern and the} \textcolor{black}{\emph{QFT}}
\textcolor{black}{one, the following (average) pattern matching metric\begin{equation}
\Gamma\triangleq\frac{1}{R}\sum_{r=1}^{R}\frac{\sum_{m=0}^{M-1}\left|\hat{P}_{m}-\hat{\wp}_{m}^{(r)}\right|}{\sum_{m=0}^{M-1}\hat{P}_{m}}\label{eq:_pattern.matching}\end{equation}
has been used, $\hat{\wp}_{m}^{(r)}$ being the normalized probability
of measuring the $m$-th ($m=0,...,M-1$) output quantum states at
the $r$-th ($1\le r\le R$) run of the} \textcolor{black}{\emph{QC}}
\textcolor{black}{process. While one should expected a more faithful
pattern matching for higher} \textcolor{black}{\emph{SLL}} \textcolor{black}{(Fig.
5), the behavior of the plots in Fig. 6 proves the opposite since
the $\Gamma$ value gets worse as} \textcolor{black}{\emph{SLL}} \textcolor{black}{increases.
Such an (apparently) contradictory outcome can be explained by separately
analyzing the mainlobe (}\textcolor{black}{\emph{ML}}\textcolor{black}{)\begin{equation}
\Gamma_{ML}=\frac{1}{R}\sum_{r=1}^{R}\frac{\sum_{m=\chi_{1}+1}^{\chi_{2}-1}\left|\hat{P}_{m}-\hat{\wp}_{m}^{(r)}\right|}{\sum_{m=0}^{M-1}\hat{P}_{m}}\label{eq:_pattern.matching.ML}\end{equation}
and the sidelobe (}\textcolor{black}{\emph{SL}}\textcolor{black}{)
\begin{equation}
\Gamma_{SL}=\frac{1}{R}\sum_{r=1}^{R}\frac{\sum_{m=-1}^{\chi_{1}}\left|\hat{P}_{m}-\hat{\wp}_{m}^{(r)}\right|+\sum_{m=\chi_{2}}^{1}\left|\hat{P}_{m}-\hat{\wp}_{m}^{(r)}\right|}{\sum_{m=0}^{M-1}\hat{P}_{m}}\label{eq:_pattern.matching.SL}\end{equation}
contributions to the pattern matching metric (\ref{eq:_pattern.matching}),
$\chi_{1}$ and $\chi_{2}$ ($\chi_{1}$, $\chi_{2}$ $\in$ $\left[0,\, M-1\right]$)
being the indexes of the pattern samples in the angular positions
closer to the nulls on the left and right of the mainlobe, respectively,
while the same normalization factor of (\ref{eq:_pattern.matching})
has been kept to fulfil the condition $\Gamma=\Gamma_{ML}+\Gamma_{SL}$.}

\noindent \textcolor{black}{The dashed-line plots of $\Gamma_{ML}$
and $\Gamma_{SL}$ in Fig. 6 indicate that the mismatch is mainly
in the} \textcolor{black}{\emph{SL}} \textcolor{black}{region (i.e.,
$\Gamma_{SL}>\Gamma_{ML}$) and the $\Gamma_{SL}$ value is greater
when the} \textcolor{black}{\emph{SLL}} \textcolor{black}{of the reference
pattern reduces (i.e., $\left.\Gamma_{SL}\right\rfloor _{SLL=-25\,[dB]}$
$>$ $\left.\Gamma_{SL}\right\rfloor _{SLL=-20\,[dB]}$ $>$ $\left.\Gamma_{SL}\right\rfloor _{SLL=-15\,[dB]}$)
according to the conclusions drawn from Fig. 3(}\textcolor{black}{\emph{b}}\textcolor{black}{)
and Figs. 5(}\textcolor{black}{\emph{a}}\textcolor{black}{)-5(}\textcolor{black}{\emph{b}}\textcolor{black}{).
Otherwise, the mainlobe index, $\Gamma_{ML}$,} \textcolor{black}{\emph{}}\textcolor{black}{decreases
(i.e., $\left.\Gamma_{ML}\right\rfloor _{SLL=-25\,[dB]}$ $<$ $\left.\Gamma_{ML}\right\rfloor _{SLL=-20\,[dB]}$
$<$ $\left.\Gamma_{ML}\right\rfloor _{SLL=-15\,[dB]}$) and it is
almost constant for smaller} \textcolor{black}{\emph{SLL}} \textcolor{black}{values
independently on $T$ (e.g., see $\Gamma_{ML}$ vs. $T$ when $SLL=-25$
{[}dB{]} - Fig. 6). Such a behavior strictly depends on the fact that
a fixed number of pattern samples, $M$, is used for the} \textcolor{black}{\emph{QFT}}
\textcolor{black}{computation and the number of samples entering in
the mainlobe, which is proportional to the so-called first-null beam
width (}\textcolor{black}{\emph{FNBW}}\textcolor{black}{), grows when
lowering the} \textcolor{black}{\emph{SLL}}\textcolor{black}{. Therefore,
patterns with wider} \textcolor{black}{\emph{FNBW}} \textcolor{black}{($\to$
lower} \textcolor{black}{\emph{SLL}}\textcolor{black}{) involve more
'high-probability' quantum states over the total number $M$ and,
since such states are more carefully predicted (even with few shots)
being the most probable, it turns out that the} \textcolor{black}{\emph{QFT}}
\textcolor{black}{performance (i.e., the total pattern matching error,
$\Gamma$) improve.}

\noindent \textcolor{black}{However, one is usually more interested
in characterizing the pattern as a whole, thus having sufficient details
of the least probable states (i.e., the low pattern values), as well.
To infer the optimal trade-off $T$, so that an accurate matching
of the reference power pattern regardless of its} \textcolor{black}{\emph{SLL}}
\textcolor{black}{is yielded, the $\Gamma_{SL}$ value of the pattern
in Fig. 2 (i.e., $\Gamma_{SL}^{th}=5.8\times10^{-2}$) has been set
as quality-target threshold and the} \textcolor{black}{\emph{QFT}}
\textcolor{black}{simulations have been run by progressively increasing
the number of shots $T$ until the condition $\Gamma_{SL}\left(T\right)\leq\Gamma_{SL}^{th}$
has hold true. The accuracy threshold $\Gamma_{SL}^{th}$ has been
reached after $\left.T\right\rfloor _{SLL=-20\,[dB]}=1.8\times M\times10^{3}$
and $\left.T\right\rfloor _{SLL=-25\,[dB]}=2.4\times M\times10^{3}$
shots, respectively (Fig. 7). Figure 8 shows the corresponding} \textcolor{black}{\emph{QFT}}
\textcolor{black}{results when predicting the reference} \textcolor{black}{\emph{DC}}
\textcolor{black}{pattern with $SLL=-20$ {[}dB{]} {[}Fig. 8(}\textcolor{black}{\emph{a}}\textcolor{black}{){]}
and $SLL=-25$ {[}dB{]} {[}Fig. 8(}\textcolor{black}{\emph{b}}\textcolor{black}{){]}.
As expected, there is a more faithful fitting of the reference pattern
in the whole angular range and, in particular, within the sidelobe
regions {[}e.g., Fig. 8(}\textcolor{black}{\emph{a}}\textcolor{black}{)
vs. Fig. 5(}\textcolor{black}{\emph{a}}\textcolor{black}{) and Fig.
8(}\textcolor{black}{\emph{b}}\textcolor{black}{) vs. Fig. 5(}\textcolor{black}{\emph{b}}\textcolor{black}{){]}.}

\noindent \textcolor{black}{While previous examples dealt with reference
patterns with equi-ripple sidelobes from} \textcolor{black}{\emph{DC}}
\textcolor{black}{distributions, the performance of the} \textcolor{black}{\emph{QC}}\textcolor{black}{-based
method have been assessed next for a $d=\frac{\lambda}{2}$-spaced
array with the $N=16$ Taylor excitations in Tab. I affording a power
pattern with decreasing sidelobe peaks when moving far from the mainlobe}
\textcolor{black}{\emph{}}\textcolor{black}{(i.e.,} \textcolor{black}{\emph{SLL}}\textcolor{black}{=$-15$
{[}dB{]} and $\bar{n}=4$). The} \textcolor{black}{\emph{QFT}} \textcolor{black}{process
has been applied by varying the number of shots from $T=M\times2$
up to $T=M\times10^{3}$ and repeating each test $R=20$ times. The
behavior of the pattern matching indexes is shown in Fig. 9(}\textcolor{black}{\emph{a}}\textcolor{black}{)
and, as a representative example, Figure 9(}\textcolor{black}{\emph{b}}\textcolor{black}{)
gives the} \textcolor{black}{\emph{QFT}} \textcolor{black}{power pattern
characterized by $\Gamma_{SL}=\Gamma_{SL}^{th}$ ($\Rightarrow$ $T=M\times10^{3}$
shots) to prove that it is possible to compute a generic pattern with
the required degree of accuracy subject to a proper choice of the
number $T$ of} \textcolor{black}{\emph{QC}} \textcolor{black}{shots.}

\noindent \textcolor{black}{The last example is concerned with the
shaped beam patterns generated by the complex sets of excitations
in Fig. 10. For both the flat-top and the cosecant-square patterns,
the results in Fig. 11 indicate that the main deviations from the
reference pattern occur in the} \textcolor{black}{\emph{SL}} \textcolor{black}{region
(i.e., $\Gamma_{SL}$ $>$ $\Gamma_{ML}$) whatever the $T$ value,
despite the main lobe occupies a non-negligible part of the whole
visible range $-1\le u\le1$. Once again the reason is that the quantum
states associated to the samples in the} \textcolor{black}{\emph{ML}}
\textcolor{black}{region (i.e., the samples of the power pattern with
higher magnitudes) have higher probability of being observed in the
measurements also for low values of $T$ (e.g., $T=M\times10$) {[}Figs.
12(}\textcolor{black}{\emph{a}}\textcolor{black}{)-12(}\textcolor{black}{\emph{b}}\textcolor{black}{){]},
while the} \textcolor{black}{\emph{SL}} \textcolor{black}{region is
better predicted when $T$ grows to $T=M\times10^{2}$ {[}Figs. 12(}\textcolor{black}{\emph{c}}\textcolor{black}{)-12(}\textcolor{black}{\emph{d}}\textcolor{black}{){]}
and $T=M\times10^{3}$ {[}Figs. 12(}\textcolor{black}{\emph{e}}\textcolor{black}{)-12(}\textcolor{black}{\emph{f}}\textcolor{black}{){]}.}

\section{\noindent \textcolor{black}{Conclusions}}

\noindent \textcolor{black}{An innovative} \textcolor{black}{\emph{QC}}\textcolor{black}{-based
method for the analysis of the power pattern of array antennas has
been proposed. It exploits the} \textcolor{black}{\emph{QFT}} \textcolor{black}{algorithm
to yield a suitable accuracy in the pattern prediction with a significant
improvement of the computational efficiency with respect to classical}
\textcolor{black}{\emph{DFT}}\textcolor{black}{-based analysis technique.
A representative set of numerical examples has been reported and discussed
to provide in depth observations on the behaviour and the performance
of the proposed method.}

\noindent \textcolor{black}{To the best of the authors' knowledge,
the main innovative contributions of this paper with respect to the
state-of-the-art lie in}

\begin{itemize}
\item \noindent \textcolor{black}{the formulation of the} \textcolor{black}{\emph{PA}}
\textcolor{black}{analysis problem in the} \textcolor{black}{\emph{QC}}
\textcolor{black}{framework;}
\item \noindent \textcolor{black}{the adaptation of the} \textcolor{black}{\emph{QFT}}
\textcolor{black}{algorithm for the computation of the} \textcolor{black}{\emph{PA}}
\textcolor{black}{power pattern and the exploitation of the relationship
between the probability values of the output quantum states and the
samples of the power pattern;}
\item \textcolor{black}{the exploitation of the quantum parallelism to yield
an exponential acceleration in the computation of the power pattern
with respect to the classical} \textcolor{black}{\emph{DFT}} \textcolor{black}{method;}
\item \textcolor{black}{the study of the dependence of the pattern prediction
accuracy of the proposed} \textcolor{black}{\emph{QC}}\textcolor{black}{-based
method on the number of shots/observations to give to the interested
readers some useful guidelines to yield a reliable and effective prediction
of the power pattern on the basis of the user needs.}
\end{itemize}
\noindent Future works, beyond the scope of this paper, will be devoted
to implement hybrid algorithms where only computationally-intensive
tasks are delegated to a quantum computer, while leaving other data
processing operations to a classical computer.

\section*{\noindent \textcolor{black}{Acknowledgements}}

\noindent This work benefited from the networking activities carried
out within the Project CYBER-PHYSICAL ELECTROMAGNETIC VISION: Context-Aware
Electromagnetic Sensing and Smart Reaction (EMvisioning) (Grant no.
2017HZJXSZ) funded by the Italian Ministry of Education, University,
and Research under the PRIN2017 Program (CUP: E64I19002530001). Moreover,
it benefited from the networking activities carried out within the
Project SPEED (Grant No. 61721001) funded by National Science Foundation
of China under the Chang-Jiang Visiting Professorship Program, the
Project 'Inversion Design Method of Structural Factors of Conformal
Load-bearing Antenna Structure based on Desired EM Performance Interval'
(Grant no. 2017HZJXSZ) funded by the National Natural Science Foundation
of China, and the Project 'Research on Uncertainty Factors and Propagation
Mechanism of Conformal Loab-bearing Antenna Structure' (Grant No.
2021JZD-003) funded by the Department of Science and Technology of
Shaanxi Province within the Program Natural Science Basic Research
Plan in Shaanxi Province. \textcolor{black}{A. Massa wishes to thank
E. Vico for her never-ending inspiration, support, guidance, and help.}

\newpage
\section*{\textcolor{black}{FIGURE CAPTIONS}}

\begin{itemize}
\item \textbf{\textcolor{black}{Figure 1.}} \textcolor{black}{Sketch of
the array architecture}
\item \textbf{\textcolor{black}{Figure 2.}} \textcolor{black}{\emph{Numerical
Assessment}} \textcolor{black}{($N=16$, $d=\frac{\lambda}{2}$,}
\textcolor{black}{\emph{DC}} \textcolor{black}{excitations, $SLL=-15$
{[}dB{]}, $M=1024$, $T=M\times10^{3}$) - Power patterns and resolution
threshold.}
\item \textbf{\textcolor{black}{Figure 3.}} \textcolor{black}{\emph{Numerical
Assessment}} \textcolor{black}{($N=16$, $d=\frac{\lambda}{2}$,}
\textcolor{black}{\emph{DC}} \textcolor{black}{excitations, $SLL=-15$
{[}dB{]}, $M=1024$) - Power patterns and resolution threshold when
(}\textcolor{black}{\emph{a}}\textcolor{black}{) $T=M\times8$, (}\textcolor{black}{\emph{b}}\textcolor{black}{)
$T=M\times20$, (}\textcolor{black}{\emph{c}}\textcolor{black}{) $T=M\times40$,
and (}\textcolor{black}{\emph{d}}\textcolor{black}{) $T=M\times80$.}
\item \textbf{\textcolor{black}{Figure 4.}} \textcolor{black}{\emph{Numerical
Assessment}} \textcolor{black}{($N=16$, $d=\frac{\lambda}{2}$,}
\textcolor{black}{\emph{DC}} \textcolor{black}{excitations, $SLL=-15$
{[}dB{]}, $M=1024$) - Behavior of the} \textcolor{black}{\emph{QFT}}
\textcolor{black}{resolution threshold statistics versus the number
of shots $T$ {[}average value - (solid line) and interval values
within the upper/lower bounds (shaded region){]}.}
\item \textbf{\textcolor{black}{Figure 5.}} \textcolor{black}{\emph{Numerical
Assessment}} \textcolor{black}{($N=16$, $d=\frac{\lambda}{2}$,}
\textcolor{black}{\emph{DC}} \textcolor{black}{excitations, $M=1024$,
$T=M\times80$) - Power patterns when (}\textcolor{black}{\emph{a}}\textcolor{black}{)
$SLL=-20$ {[}dB{]} and (}\textcolor{black}{\emph{b}}\textcolor{black}{)
$SLL=-25$ {[}dB{]}.}
\item \textbf{\textcolor{black}{Figure 6.}} \textcolor{black}{\emph{Numerical}}
\textcolor{black}{}\textcolor{black}{\emph{Assessment}} \textcolor{black}{($N=16$,
$d=\frac{\lambda}{2}$,} \textcolor{black}{\emph{DC}} \textcolor{black}{excitations,
$M=1024$) - Behavior of the pattern matching metrics versus the number
of shots $T$.}
\item \textbf{\textcolor{black}{Figure 7.}} \textcolor{black}{\emph{Numerical}}
\textcolor{black}{}\textcolor{black}{\emph{Assessment}} \textcolor{black}{($N=16$,
$d=\frac{\lambda}{2}$,} \textcolor{black}{\emph{DC}} \textcolor{black}{excitations,
$M=1024$) - Behaviour of the} \textcolor{black}{\emph{SL}} \textcolor{black}{contribution
of the pattern matching error, $\Gamma_{SL}$, versus the number of
shots $T$.}
\item \textbf{\textcolor{black}{Figure 8.}} \textcolor{black}{\emph{Numerical
Assessment}} \textcolor{black}{($N=16$, $d=\frac{\lambda}{2}$,}
\textcolor{black}{\emph{DC}} \textcolor{black}{excitations, $M=1024$)
- Power patterns when (}\textcolor{black}{\emph{a}}\textcolor{black}{)
$SLL=-20$ {[}dB{]} and $T=1.8\times10^{3}\times M$ and (}\textcolor{black}{\emph{b}}\textcolor{black}{)
$SLL=-25$ {[}dB{]} and $T=2.4\times10^{3}\times M$.}
\item \textbf{\textcolor{black}{Figure 9.}} \textcolor{black}{\emph{Numerica}}\textcolor{black}{l}
\textcolor{black}{\emph{Assessment}} \textcolor{black}{($N=16$, $d=\frac{\lambda}{2}$,}
\textcolor{black}{\emph{Taylor}} \textcolor{black}{excitations, $M=1024$)
- Plots of (}\textcolor{black}{\emph{a}}\textcolor{black}{) pattern
matching metrics versus the number of shots $T$ and (}\textcolor{black}{\emph{b}}\textcolor{black}{)
power pattern when $T=M\times10^{3}$.}
\item \textbf{\textcolor{black}{Figure 10.}} \textcolor{black}{\emph{Numerical}}
\textcolor{black}{}\textcolor{black}{\emph{Assessment}} \textcolor{black}{($N=16$,
$d=\frac{\lambda}{2}$) - Normalized excitations of the array elements.}
\item \textbf{\textcolor{black}{Figure 11.}} \textcolor{black}{\emph{Numerica}}\textcolor{black}{l}
\textcolor{black}{\emph{Assessment}} \textcolor{black}{($N=16$, $M=1024$)
- Behavior of the pattern matching metrics versus the number of shots
$T$.}
\item \textbf{\textcolor{black}{Figure 12.}} \textcolor{black}{\emph{Numerica}}\textcolor{black}{l}
\textcolor{black}{\emph{Assessment}} \textcolor{black}{($N=16$, $M=1024$)
- Plots of the power patterns of the (}\textcolor{black}{\emph{a}}\textcolor{black}{)(}\textcolor{black}{\emph{c}}\textcolor{black}{)(}\textcolor{black}{\emph{e}}\textcolor{black}{)
flat-top and (}\textcolor{black}{\emph{b}}\textcolor{black}{)(}\textcolor{black}{\emph{d}}\textcolor{black}{)(}\textcolor{black}{\emph{f}}\textcolor{black}{)
cosecant-square beams when (}\textcolor{black}{\emph{a}}\textcolor{black}{)(}\textcolor{black}{\emph{b}}\textcolor{black}{)
$T=M\times10$, (}\textcolor{black}{\emph{c}}\textcolor{black}{)(}\textcolor{black}{\emph{d}}\textcolor{black}{)
$T=M\times10^{2}$, and (}\textcolor{black}{\emph{e}}\textcolor{black}{)(}\textcolor{black}{\emph{f}}\textcolor{black}{)
$T=M\times10^{3}$.}
\end{itemize}

\section*{\textcolor{black}{TABLE CAPTIONS}}

\begin{itemize}
\item \textbf{\textcolor{black}{Table I.}} \textcolor{black}{\emph{Numerical
Assessment}} \textcolor{black}{($N=16$,} \textcolor{black}{\emph{}}\textcolor{black}{$d=\frac{\lambda}{2}$)}
\textcolor{black}{\emph{-}} \textcolor{black}{Normalized excitations.}
\end{itemize}
\newpage
\begin{center}\textcolor{black}{~\vfill}\end{center}

\begin{center}\textcolor{black}{}\begin{tabular}{c}
\textcolor{black}{\includegraphics[%
  width=0.80\columnwidth,
  height=0.60\columnwidth,
  keepaspectratio]{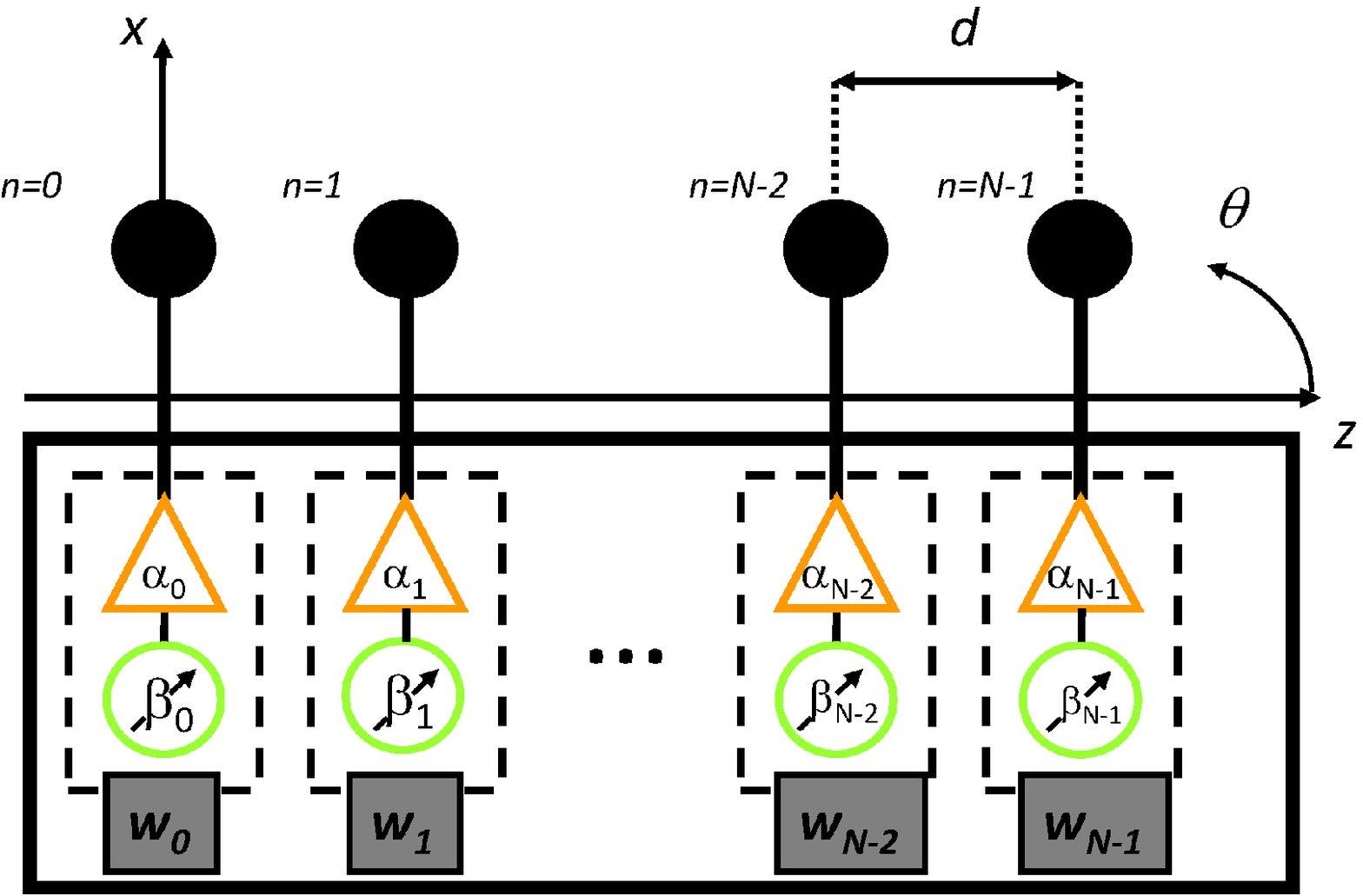}}\tabularnewline
\end{tabular}\end{center}

\begin{center}\textcolor{black}{~\vfill}\end{center}

\begin{center}\textbf{\textcolor{black}{Fig. 1 - L. Tosi et}} \textbf{\textcolor{black}{\emph{al.}}}\textbf{\textcolor{black}{,}}
\textcolor{black}{{}``Array Antenna Power Pattern Analysis ...''}\end{center}
\newpage

\begin{center}\textcolor{black}{~\vfill}\end{center}

\begin{center}\textcolor{black}{}\begin{tabular}{c}
\textcolor{black}{\includegraphics[%
  width=0.80\columnwidth,
  keepaspectratio]{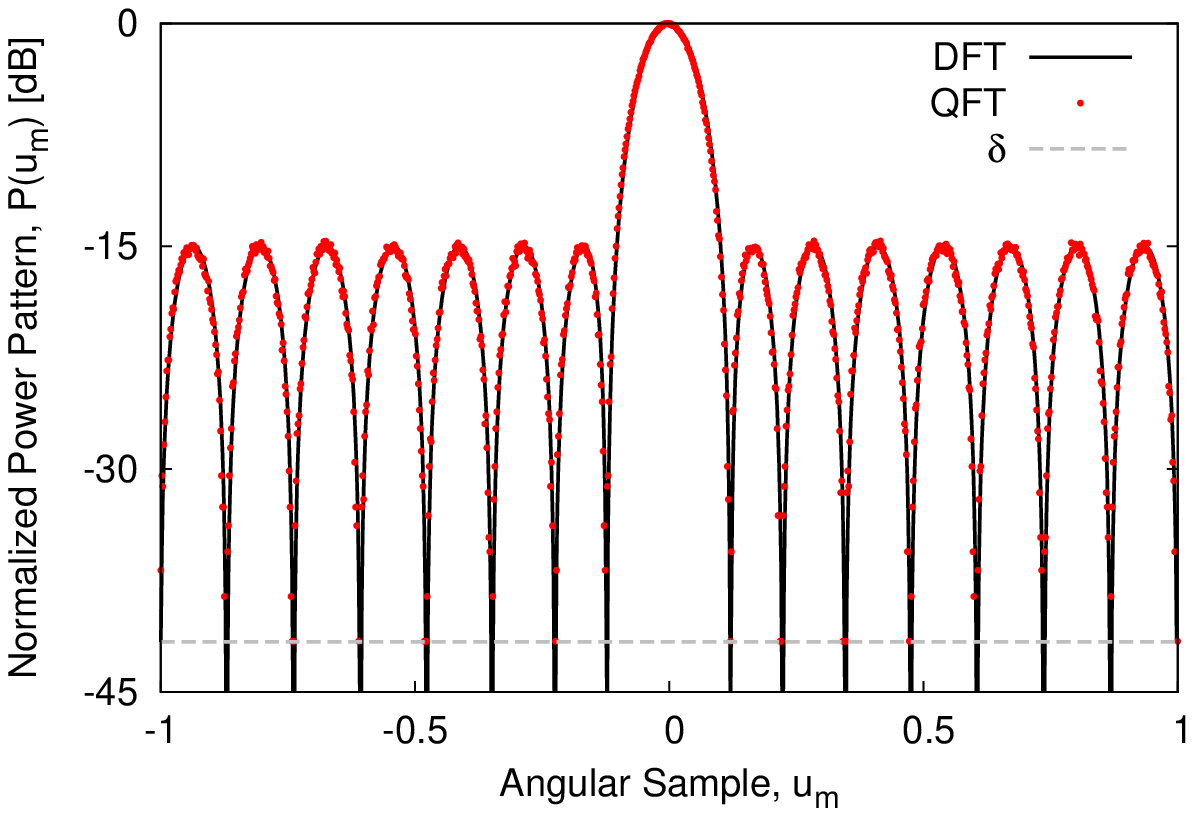}}\tabularnewline
\end{tabular}\end{center}

\begin{center}\textcolor{black}{~\vfill}\end{center}

\begin{center}\textbf{\textcolor{black}{Fig. 2 - L. Tosi et}} \textbf{\textcolor{black}{\emph{al.}}}\textbf{\textcolor{black}{,}}
\textcolor{black}{{}``Array Antenna Power Pattern Analysis ...''}\end{center}
\newpage

\begin{center}\textcolor{black}{~\vfill}\end{center}

\begin{center}\textcolor{black}{}\begin{tabular}{cc}
\textcolor{black}{\includegraphics[%
  width=0.45\columnwidth]{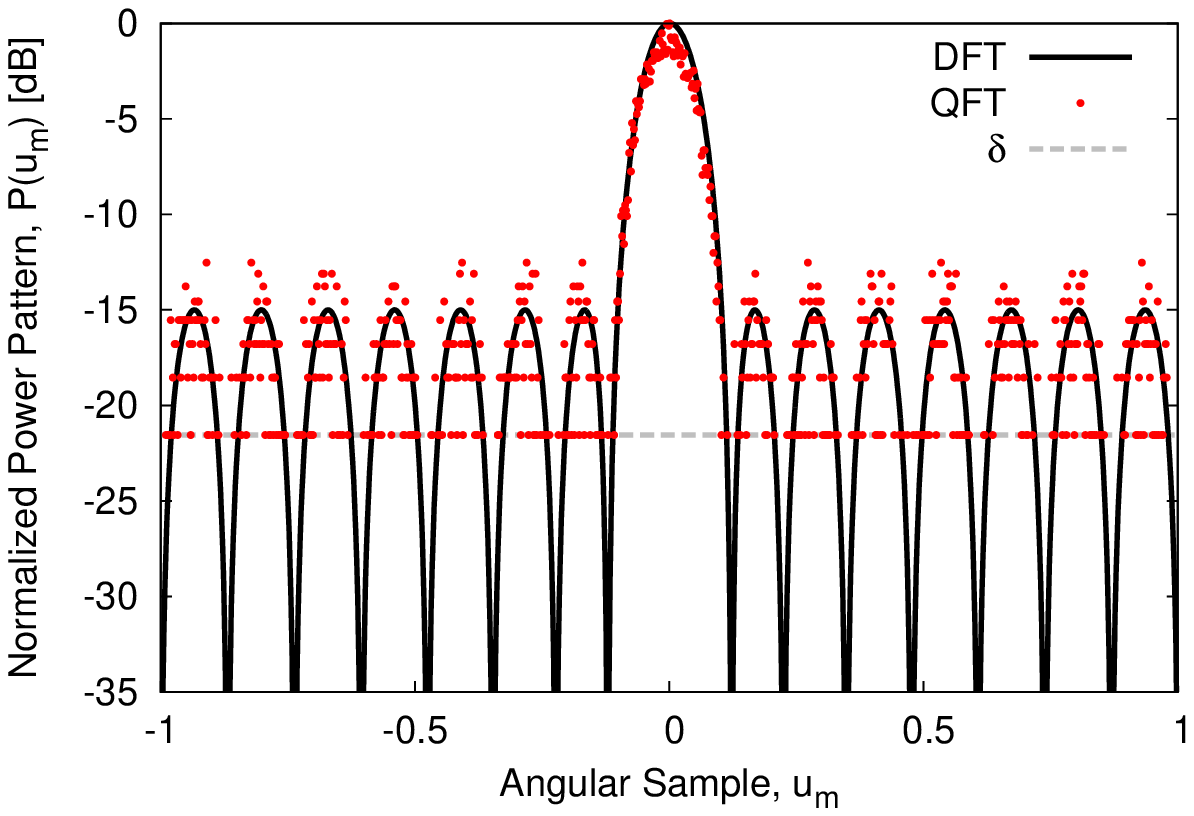}}&
\textcolor{black}{\includegraphics[%
  width=0.45\columnwidth]{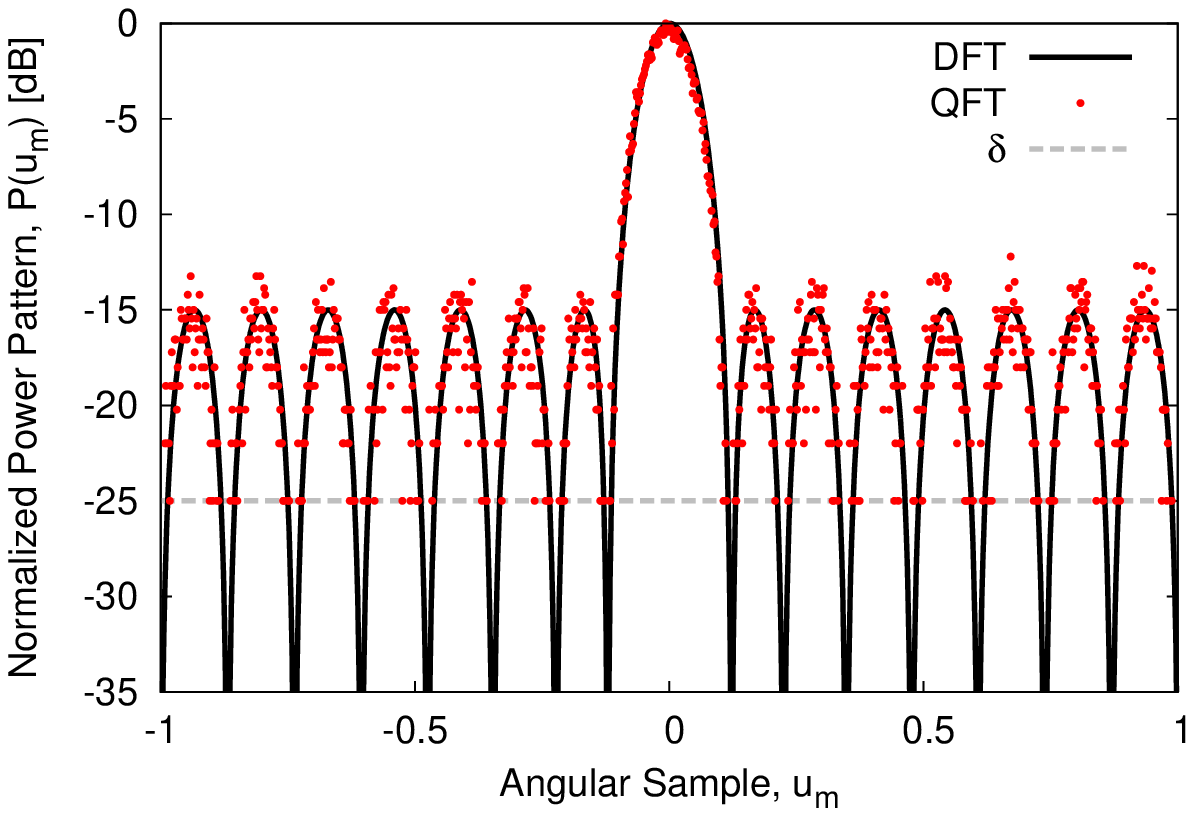}}\tabularnewline
\textcolor{black}{(}\textcolor{black}{\emph{a}}\textcolor{black}{)}&
\textcolor{black}{(}\textcolor{black}{\emph{b}}\textcolor{black}{)}\tabularnewline
\textcolor{black}{\includegraphics[%
  width=0.45\columnwidth]{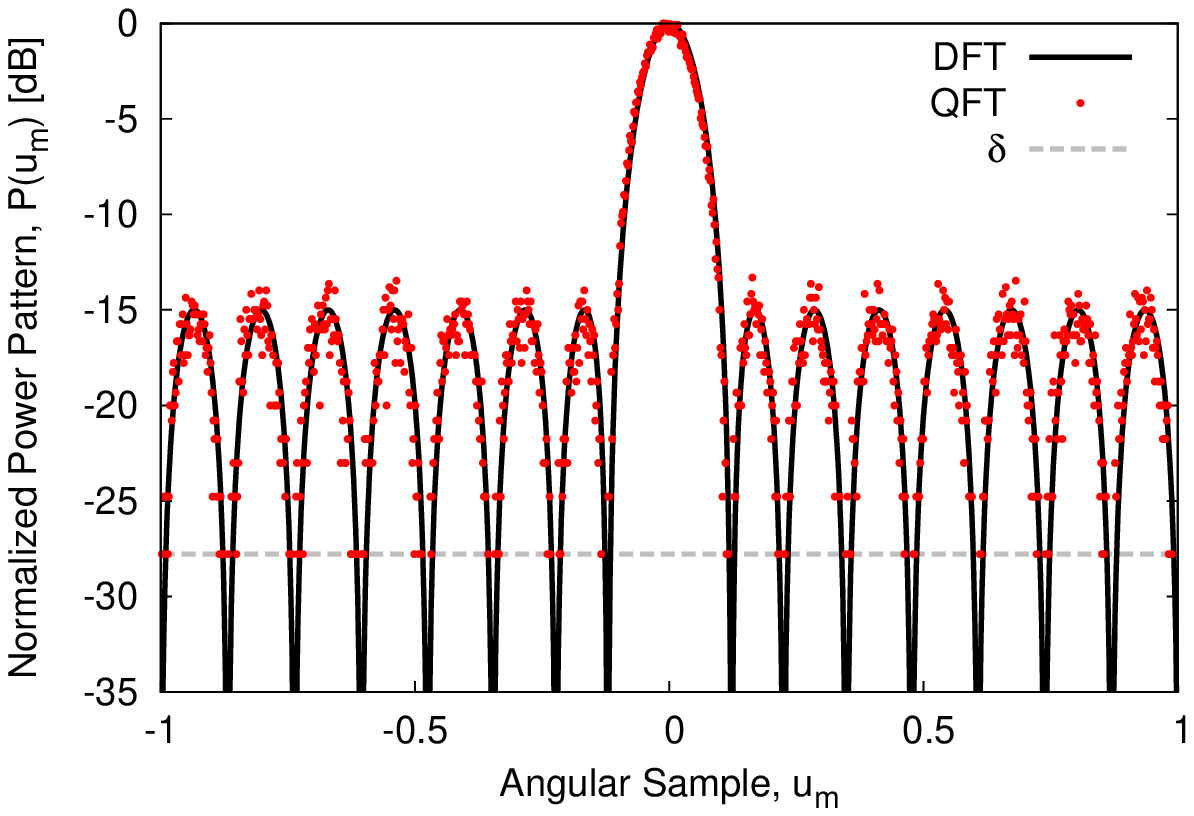}}&
\textcolor{black}{\includegraphics[%
  width=0.45\columnwidth]{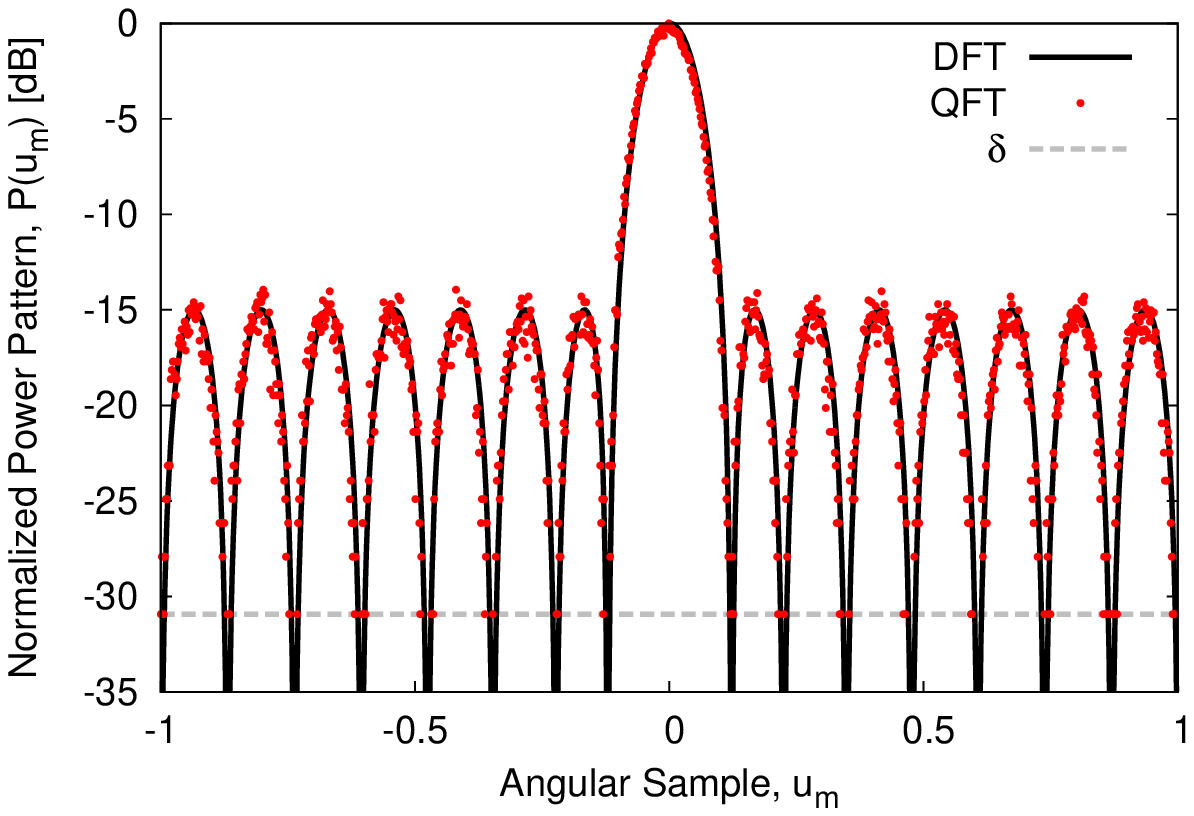}}\tabularnewline
\textcolor{black}{(}\textcolor{black}{\emph{c}}\textcolor{black}{)}&
\textcolor{black}{(}\textcolor{black}{\emph{d}}\textcolor{black}{)}\tabularnewline
\end{tabular}\end{center}

\begin{center}\textcolor{black}{~\vfill}\end{center}

\begin{center}\textbf{\textcolor{black}{Fig. 3 - L. Tosi et}} \textbf{\textcolor{black}{\emph{al.}}}\textbf{\textcolor{black}{,}}
\textcolor{black}{{}``Array Antenna Power Pattern Analysis ...''}\end{center}
\newpage

\begin{center}\textcolor{black}{~\vfill}\end{center}

\begin{center}\textcolor{black}{}\begin{tabular}{c}
\textcolor{black}{\includegraphics[%
  width=0.90\columnwidth]{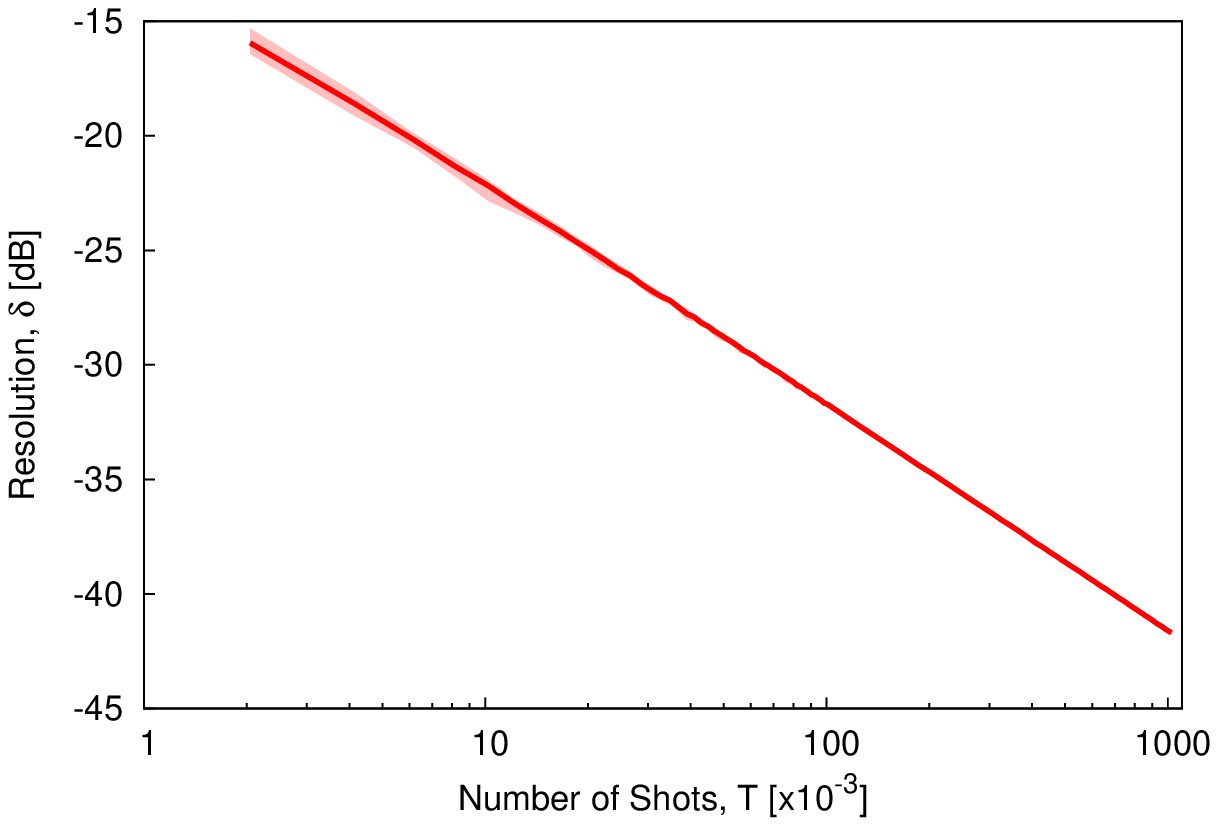}}\tabularnewline
\end{tabular}\end{center}

\begin{center}\textcolor{black}{~\vfill}\end{center}

\begin{center}\textbf{\textcolor{black}{Fig. 4 - L. Tosi et}} \textbf{\textcolor{black}{\emph{al.}}}\textbf{\textcolor{black}{,}}
\textcolor{black}{{}``Array Antenna Power Pattern Analysis ...''}\end{center}
\newpage

\begin{center}\textcolor{black}{~\vfill}\end{center}

\begin{center}\textcolor{black}{}\begin{tabular}{c}
\textcolor{black}{\includegraphics[%
  width=0.60\columnwidth]{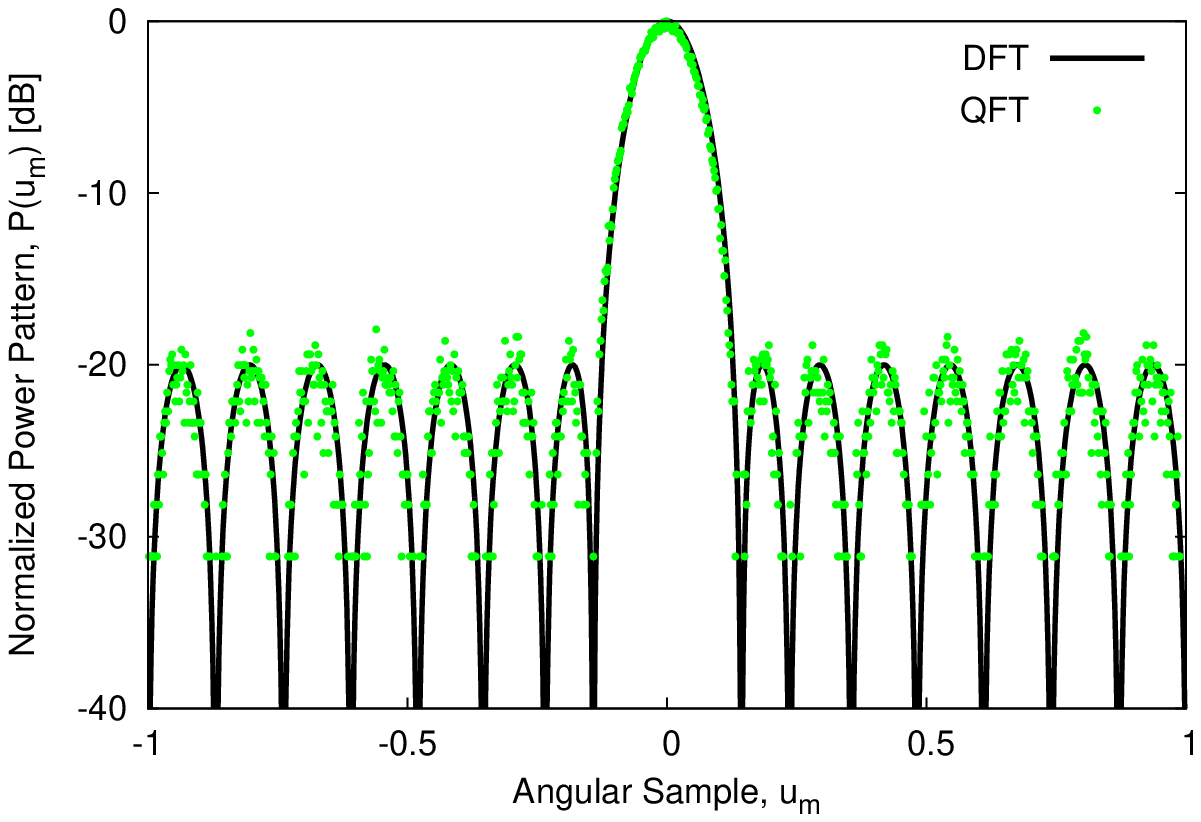}}\tabularnewline
\textcolor{black}{(}\textcolor{black}{\emph{a}}\textcolor{black}{)}\tabularnewline
\textcolor{black}{\includegraphics[%
  width=0.60\columnwidth]{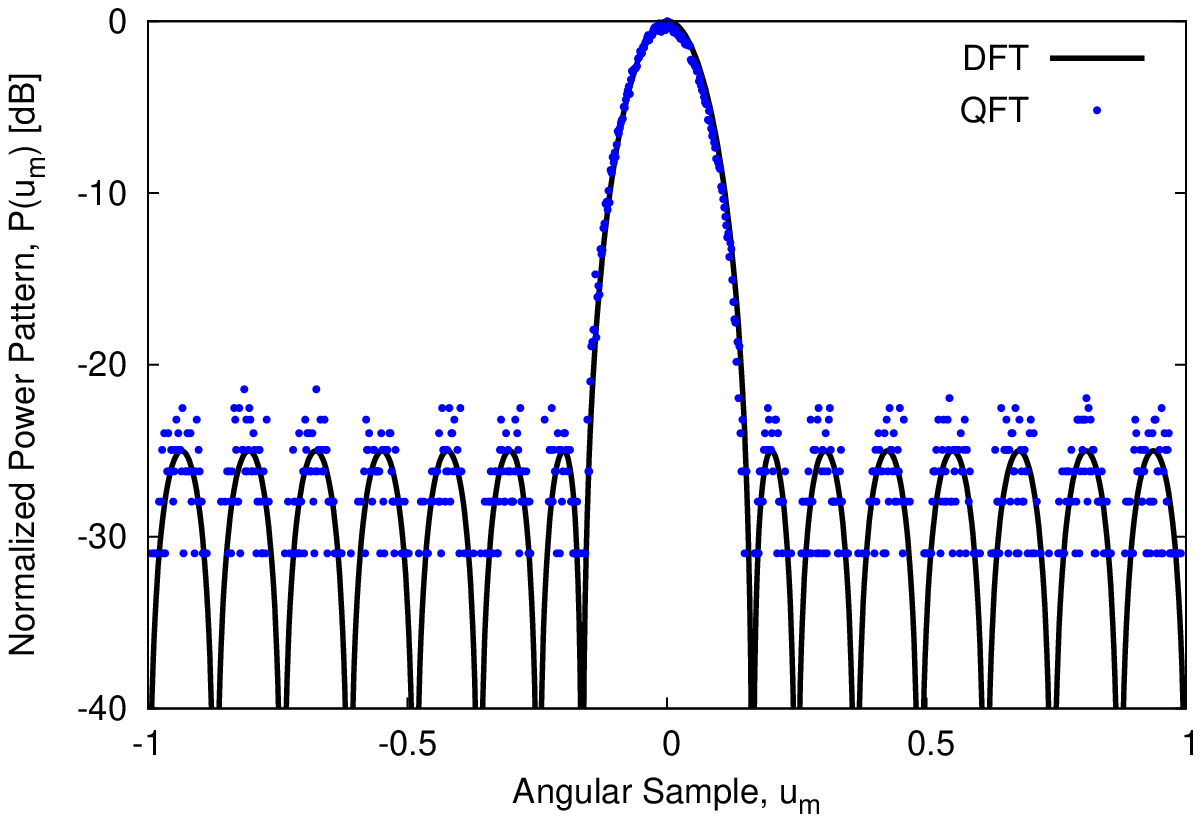}}\tabularnewline
\textcolor{black}{(}\textcolor{black}{\emph{b}}\textcolor{black}{)}\tabularnewline
\end{tabular}\end{center}

\begin{center}\textcolor{black}{~\vfill}\end{center}

\begin{center}\textbf{\textcolor{black}{Fig. 5 - L. Tosi et}} \textbf{\textcolor{black}{\emph{al.}}}\textbf{\textcolor{black}{,}}
\textcolor{black}{{}``Array Antenna Power Pattern Analysis ...''}\end{center}
\newpage

\begin{center}\textcolor{black}{~\vfill}\end{center}

\begin{center}\textcolor{black}{}\begin{tabular}{c}
\textcolor{black}{\includegraphics[%
  width=0.80\columnwidth,
  keepaspectratio]{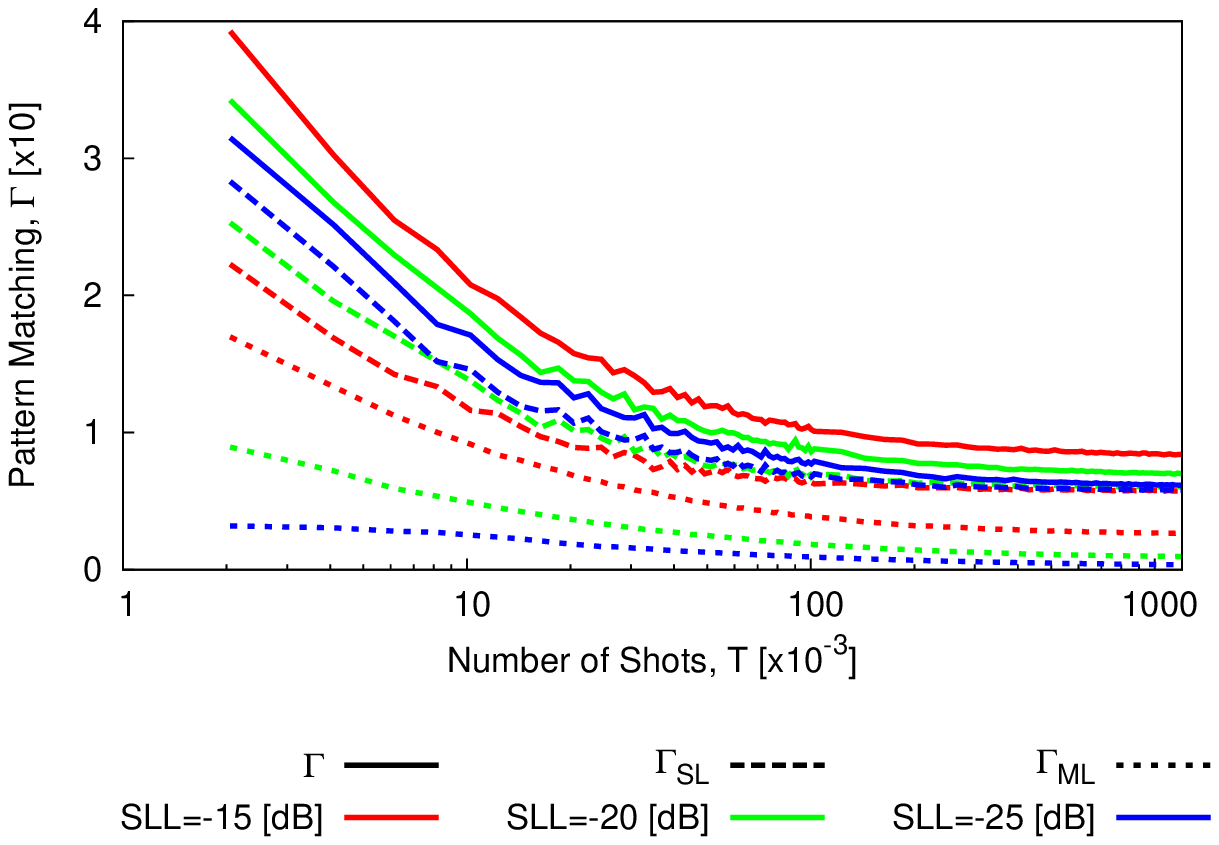}}\tabularnewline
\end{tabular}\end{center}

\begin{center}\textcolor{black}{~\vfill}\end{center}

\begin{center}\textbf{\textcolor{black}{Fig. 6 - L. Tosi et}} \textbf{\textcolor{black}{\emph{al.}}}\textbf{\textcolor{black}{,}}
\textcolor{black}{{}``Array Antenna Power Pattern Analysis ...''}\end{center}
\newpage

\textcolor{black}{~\vfill}

\begin{center}\textcolor{black}{}\begin{tabular}{c}
\textcolor{black}{\includegraphics[%
  width=0.80\columnwidth,
  keepaspectratio]{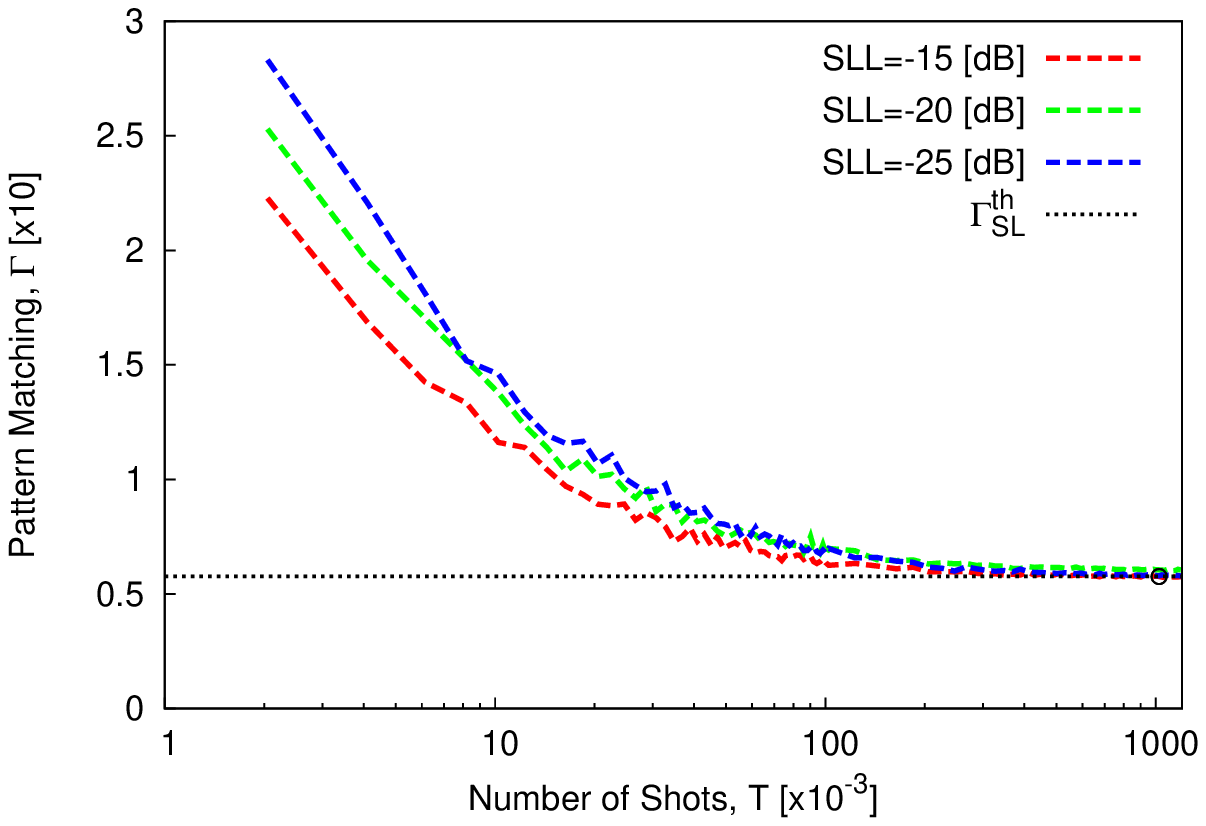}}\tabularnewline
\end{tabular}\end{center}

\begin{center}\textcolor{black}{~\vfill}\end{center}

\begin{center}\textbf{\textcolor{black}{Fig. 7 - L. Tosi et}} \textbf{\textcolor{black}{\emph{al.}}}\textbf{\textcolor{black}{,}}
\textcolor{black}{{}``Array Antenna Power Pattern Analysis ...''}\end{center}
\newpage

\textcolor{black}{~\vfill}

\begin{center}\textcolor{black}{}\begin{tabular}{c}
\textcolor{black}{\includegraphics[%
  width=0.60\columnwidth,
  keepaspectratio]{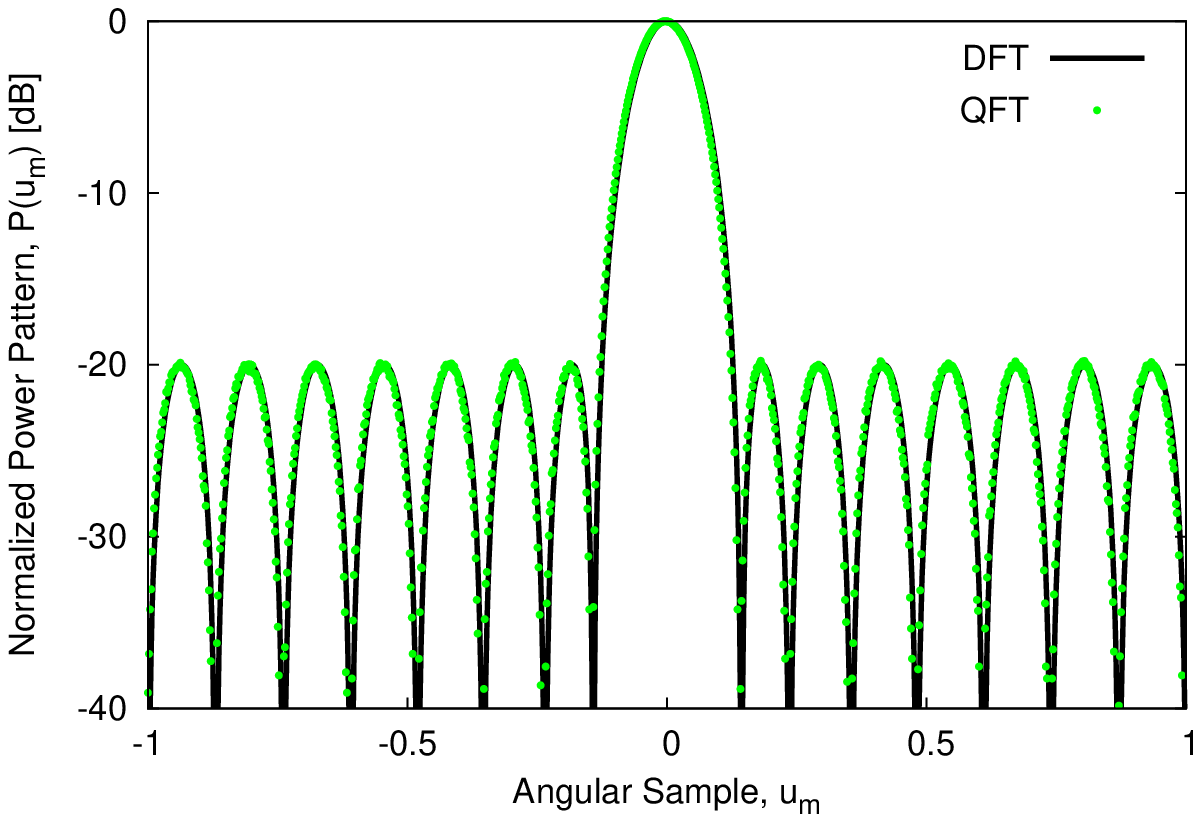}}\tabularnewline
\textcolor{black}{(}\textcolor{black}{\emph{a}}\textcolor{black}{)}\tabularnewline
\textcolor{black}{\includegraphics[%
  width=0.60\columnwidth,
  keepaspectratio]{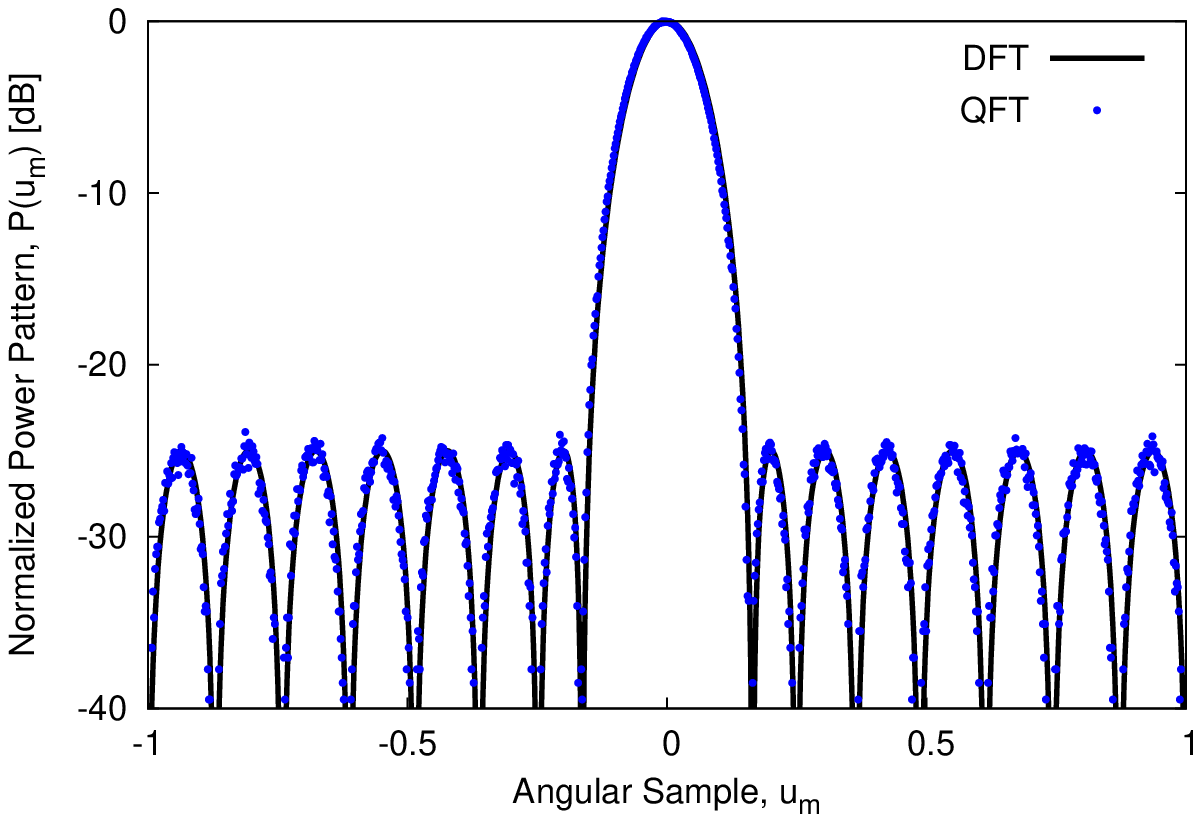}}\tabularnewline
\textcolor{black}{(}\textcolor{black}{\emph{b}}\textcolor{black}{)}\tabularnewline
\end{tabular}\end{center}

\begin{center}\textcolor{black}{~\vfill}\end{center}

\begin{center}\textbf{\textcolor{black}{Fig. 8 - L. Tosi et}} \textbf{\textcolor{black}{\emph{al.}}}\textbf{\textcolor{black}{,}}
\textcolor{black}{{}``Array Antenna Power Pattern Analysis ...''}\end{center}
\newpage

\begin{center}\textcolor{black}{~\vfill}\end{center}

\begin{center}\textcolor{black}{}\begin{tabular}{c}
\textcolor{black}{\includegraphics[%
  width=0.60\columnwidth,
  keepaspectratio]{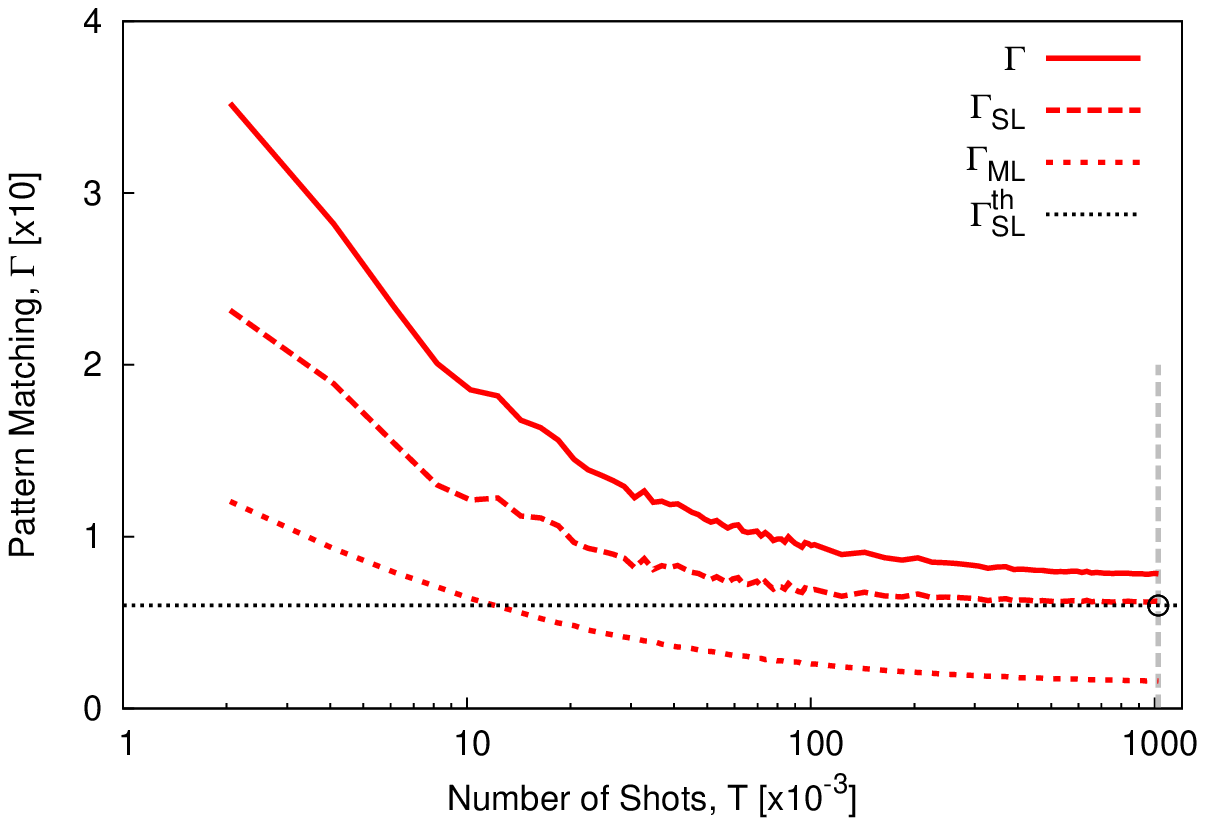}}\tabularnewline
\textcolor{black}{(}\textcolor{black}{\emph{a}}\textcolor{black}{)}\tabularnewline
\textcolor{black}{\includegraphics[%
  width=0.60\columnwidth,
  keepaspectratio]{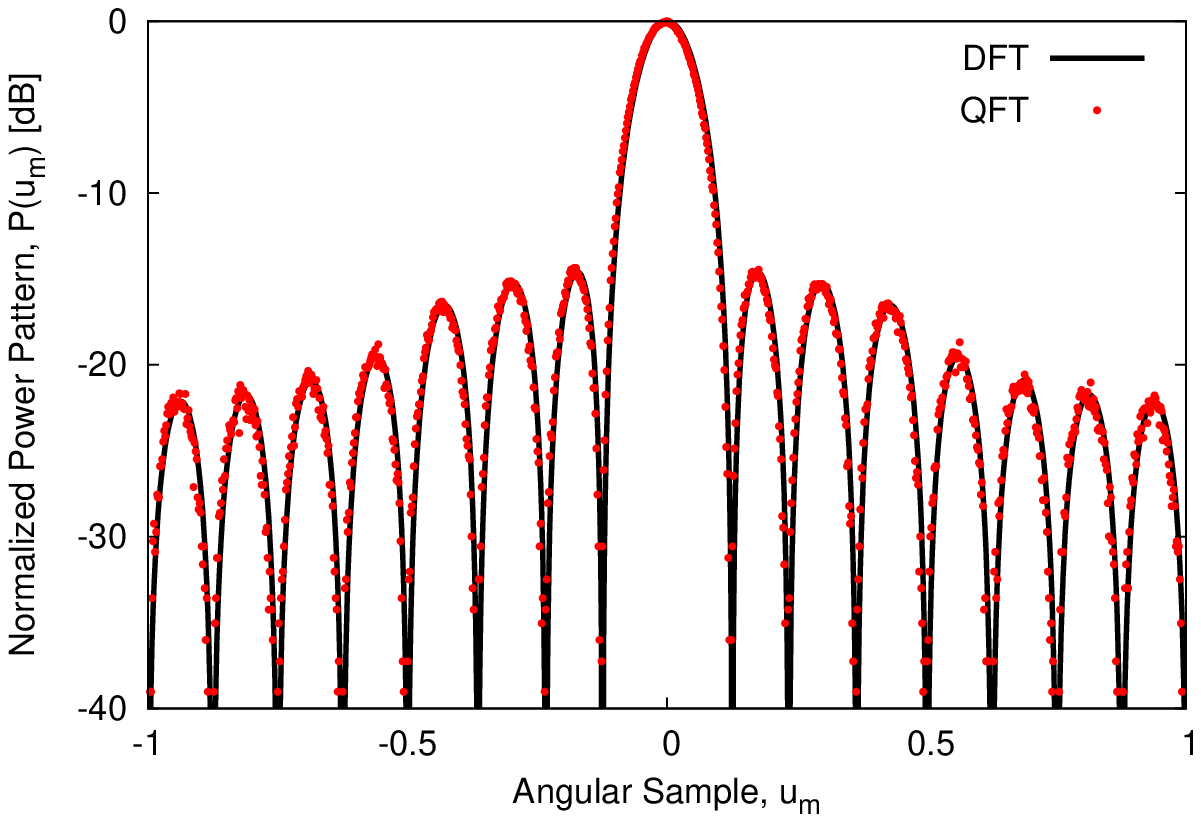}}\tabularnewline
\textcolor{black}{(}\textcolor{black}{\emph{b}}\textcolor{black}{)}\tabularnewline
\end{tabular}\end{center}

\begin{center}\textcolor{black}{~\vfill}\end{center}

\begin{center}\textbf{\textcolor{black}{Fig. 9 - L. Tosi et}} \textbf{\textcolor{black}{\emph{al.}}}\textbf{\textcolor{black}{,}}
\textcolor{black}{{}``Array Antenna Power Pattern Analysis ...''}\end{center}
\newpage

\textcolor{black}{~\vfill}

\begin{center}\textcolor{black}{}\begin{tabular}{c}
\textcolor{black}{\includegraphics[%
  width=0.70\columnwidth,
  keepaspectratio]{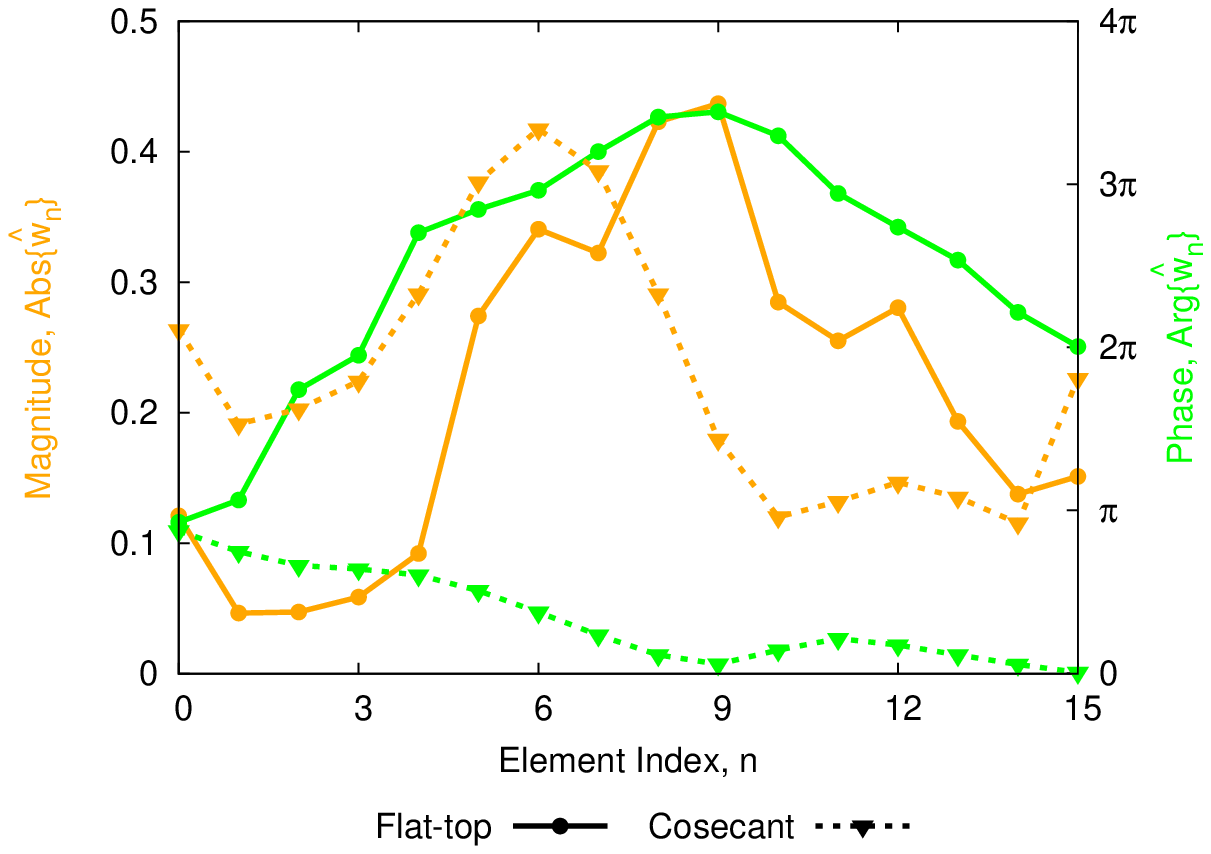}}\tabularnewline
\end{tabular}\end{center}

\begin{center}\textcolor{black}{~\vfill}\end{center}

\begin{center}\textbf{\textcolor{black}{Fig. 10 - L. Tosi et}} \textbf{\textcolor{black}{\emph{al.}}}\textbf{\textcolor{black}{,}}
\textcolor{black}{{}``Array Antenna Power Pattern Analysis ...''}\end{center}
\newpage

\textcolor{black}{~\vfill}

\begin{center}\textcolor{black}{}\begin{tabular}{c}
\textcolor{black}{\includegraphics[%
  width=0.80\columnwidth,
  keepaspectratio]{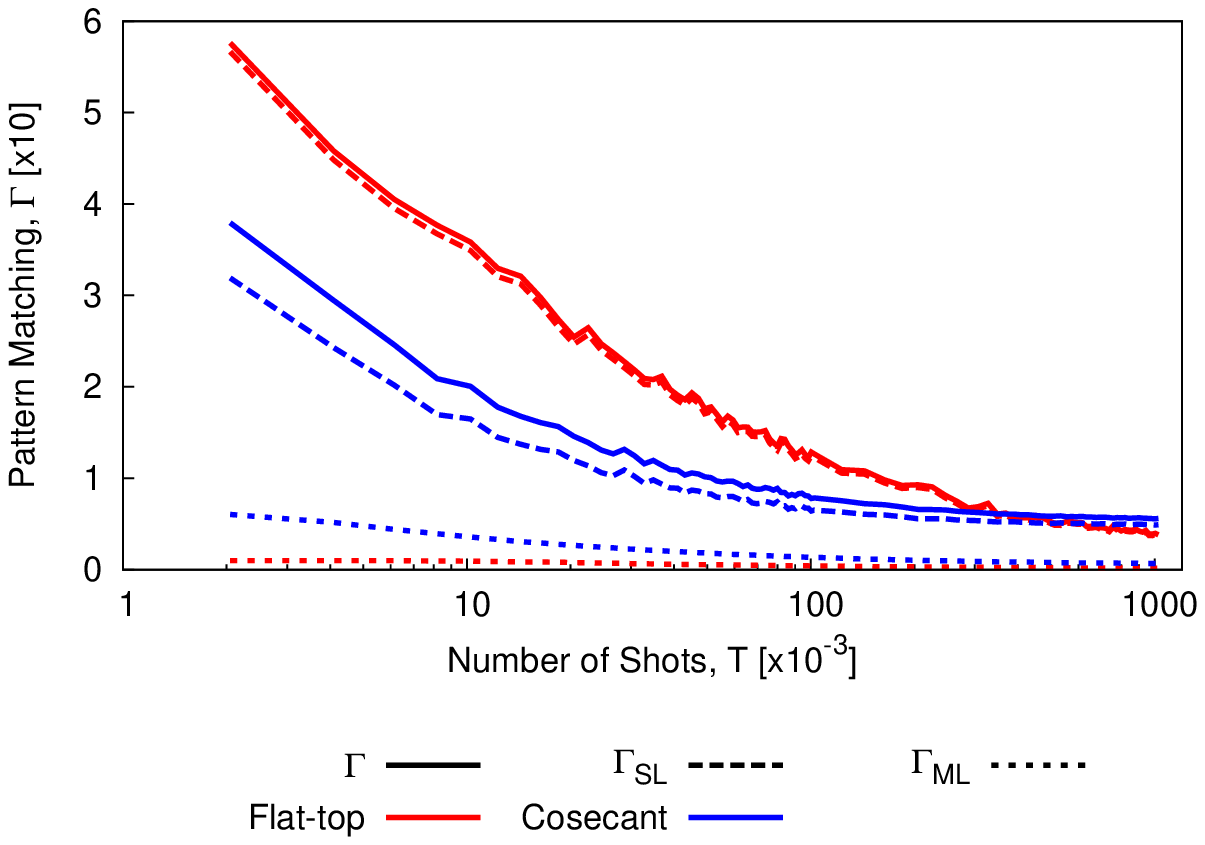}}\tabularnewline
\end{tabular}\end{center}

\begin{center}\textcolor{black}{~\vfill}\end{center}

\begin{center}\textbf{\textcolor{black}{Fig. 11 - L. Tosi et}} \textbf{\textcolor{black}{\emph{al.}}}\textbf{\textcolor{black}{,}}
\textcolor{black}{{}``Array Antenna Power Pattern Analysis ...''}\end{center}
\newpage

\textcolor{black}{~\vfill}

\begin{center}\textcolor{black}{}\begin{tabular}{cc}
\textcolor{black}{\includegraphics[%
  width=0.50\columnwidth,
  keepaspectratio]{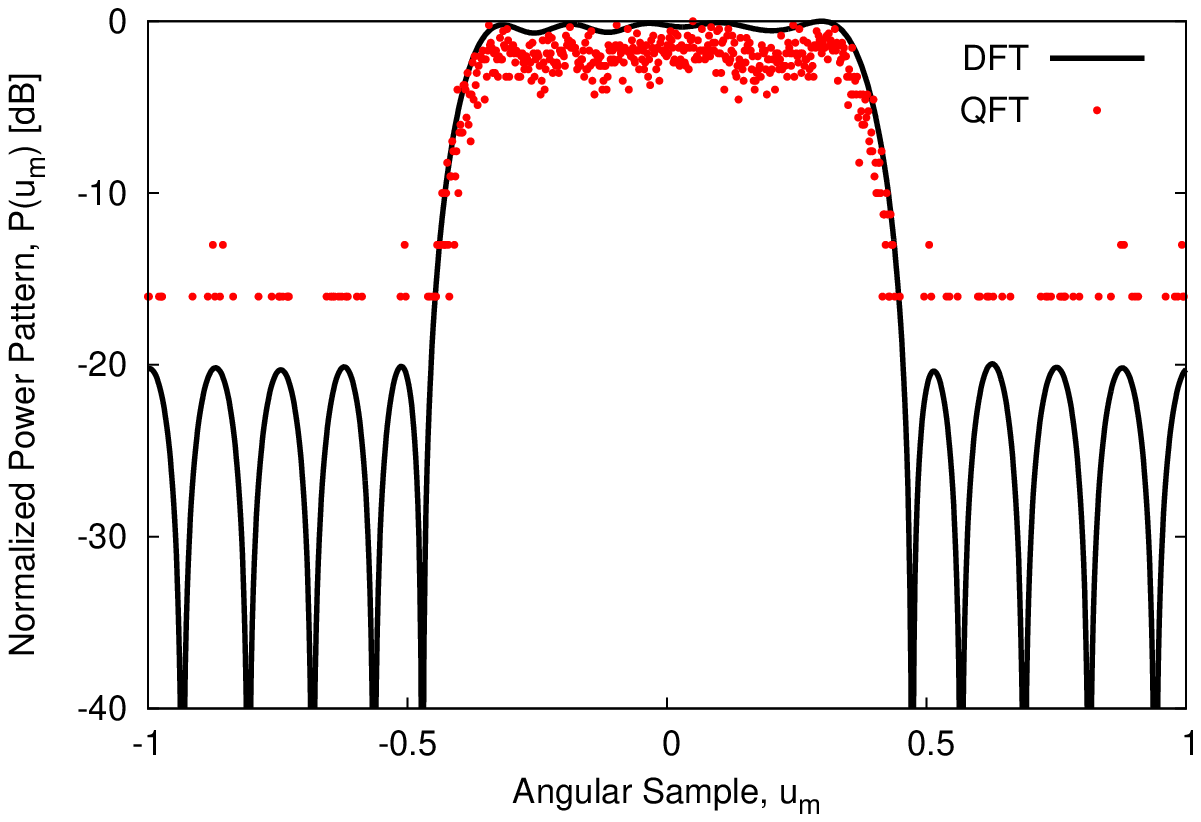}}&
\textcolor{black}{\includegraphics[%
  width=0.50\columnwidth,
  keepaspectratio]{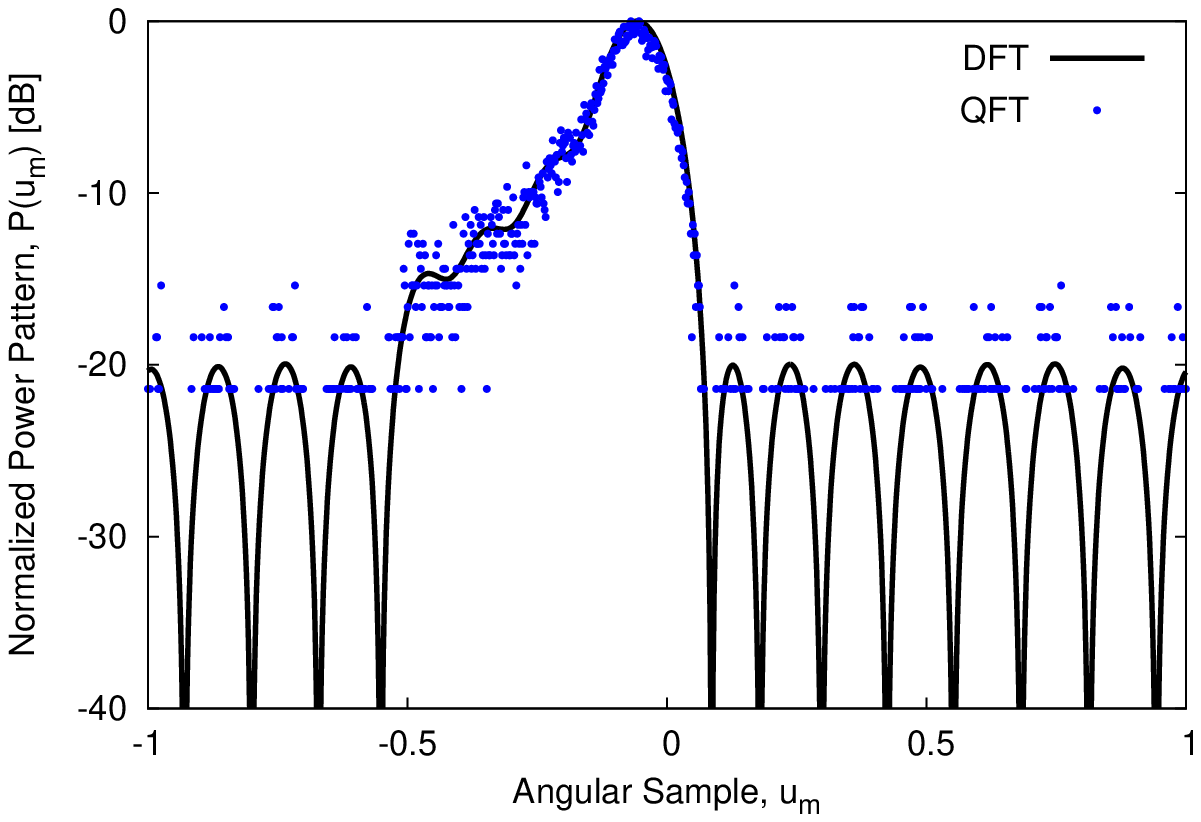}}\tabularnewline
\textcolor{black}{(}\textcolor{black}{\emph{a}}\textcolor{black}{)}&
\textcolor{black}{(}\textcolor{black}{\emph{b}}\textcolor{black}{)}\tabularnewline
\textcolor{black}{\includegraphics[%
  width=0.50\columnwidth,
  keepaspectratio]{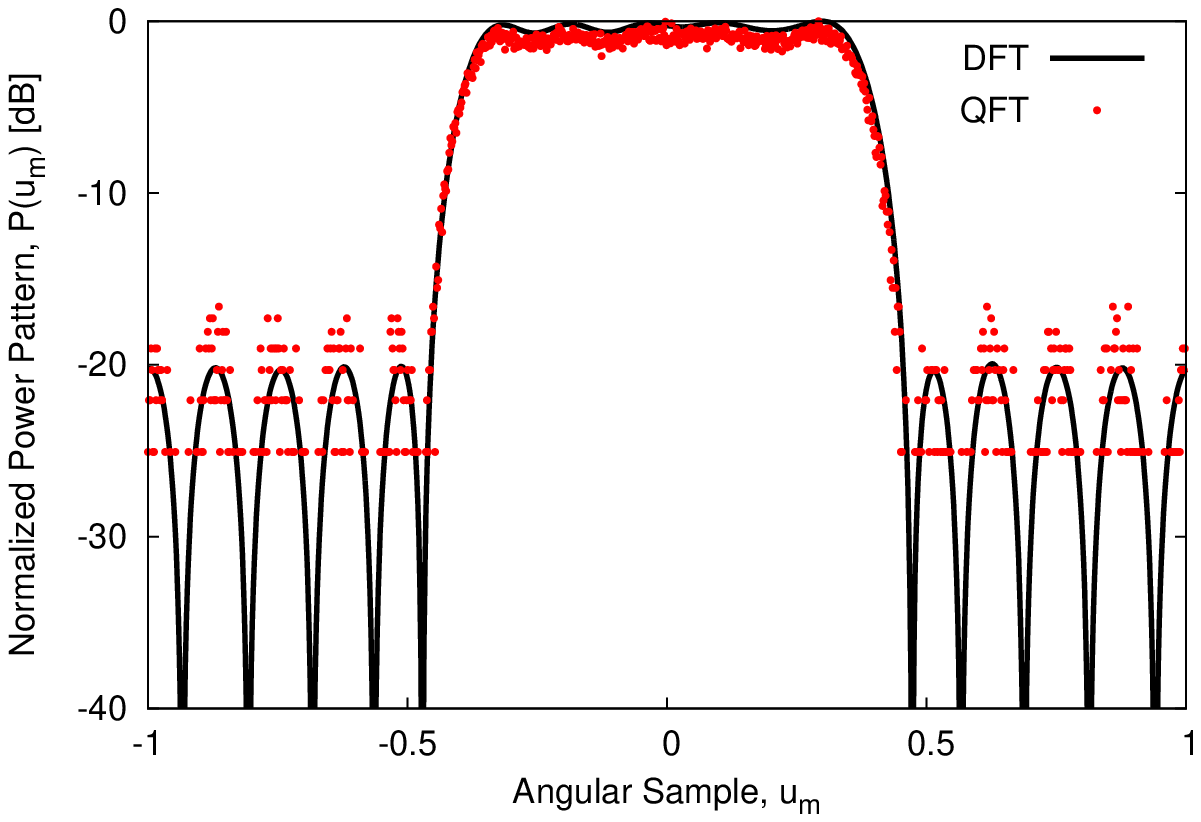}}&
\textcolor{black}{\includegraphics[%
  width=0.50\columnwidth,
  keepaspectratio]{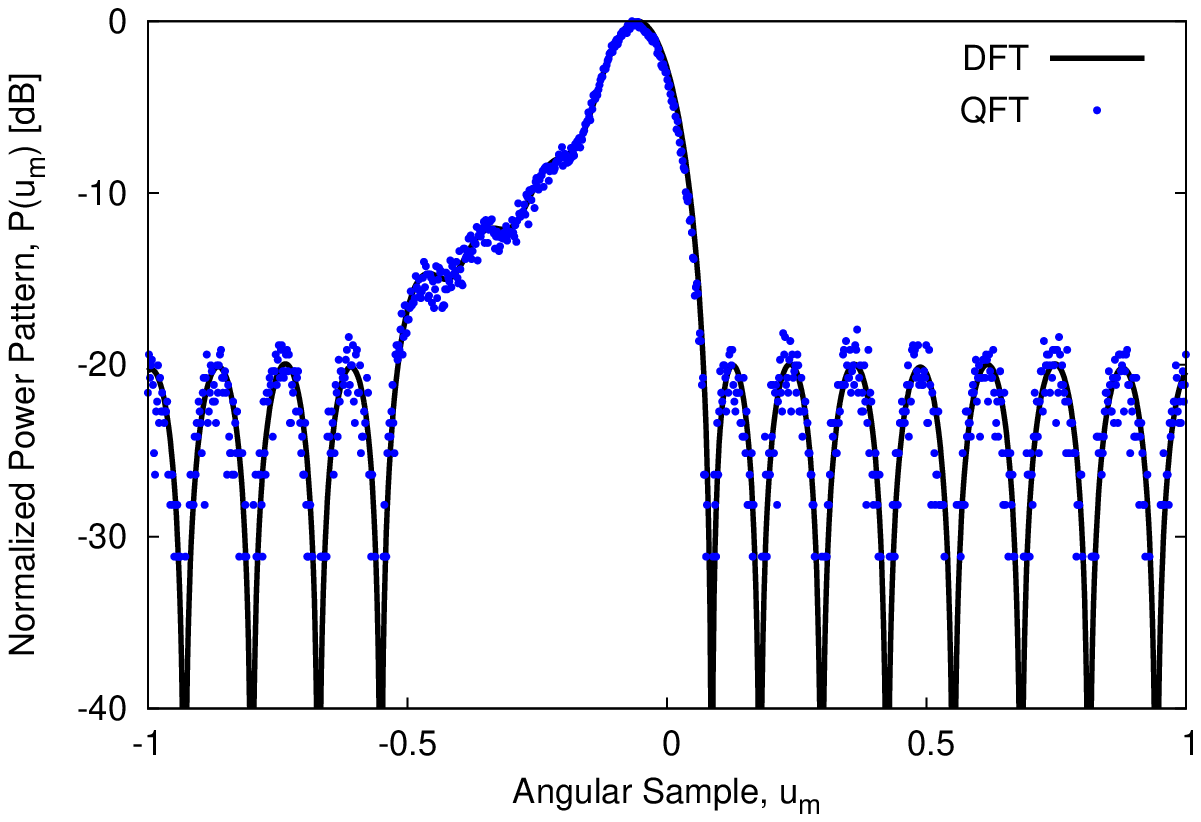}}\tabularnewline
\textcolor{black}{(}\textcolor{black}{\emph{c}}\textcolor{black}{)}&
\textcolor{black}{(}\textcolor{black}{\emph{d}}\textcolor{black}{)}\tabularnewline
\textcolor{black}{\includegraphics[%
  width=0.50\columnwidth,
  keepaspectratio]{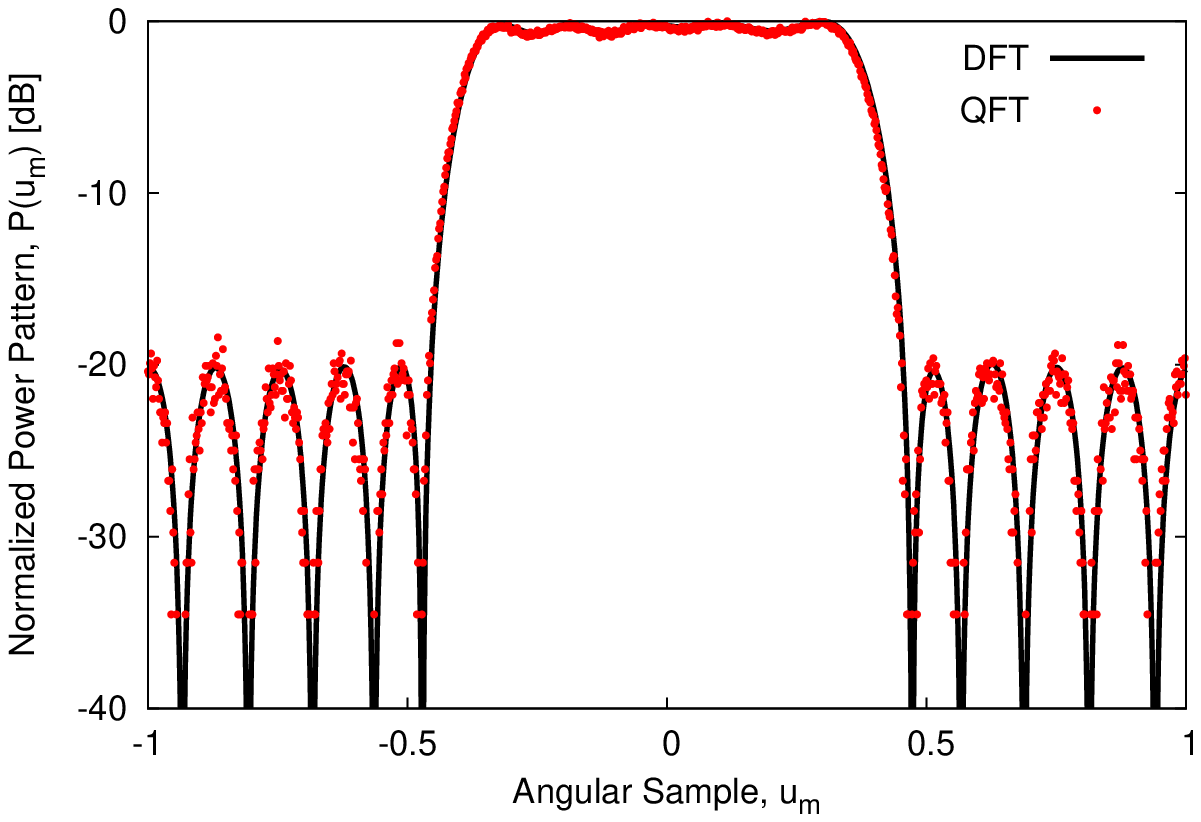}}&
\textcolor{black}{\includegraphics[%
  width=0.50\columnwidth,
  keepaspectratio]{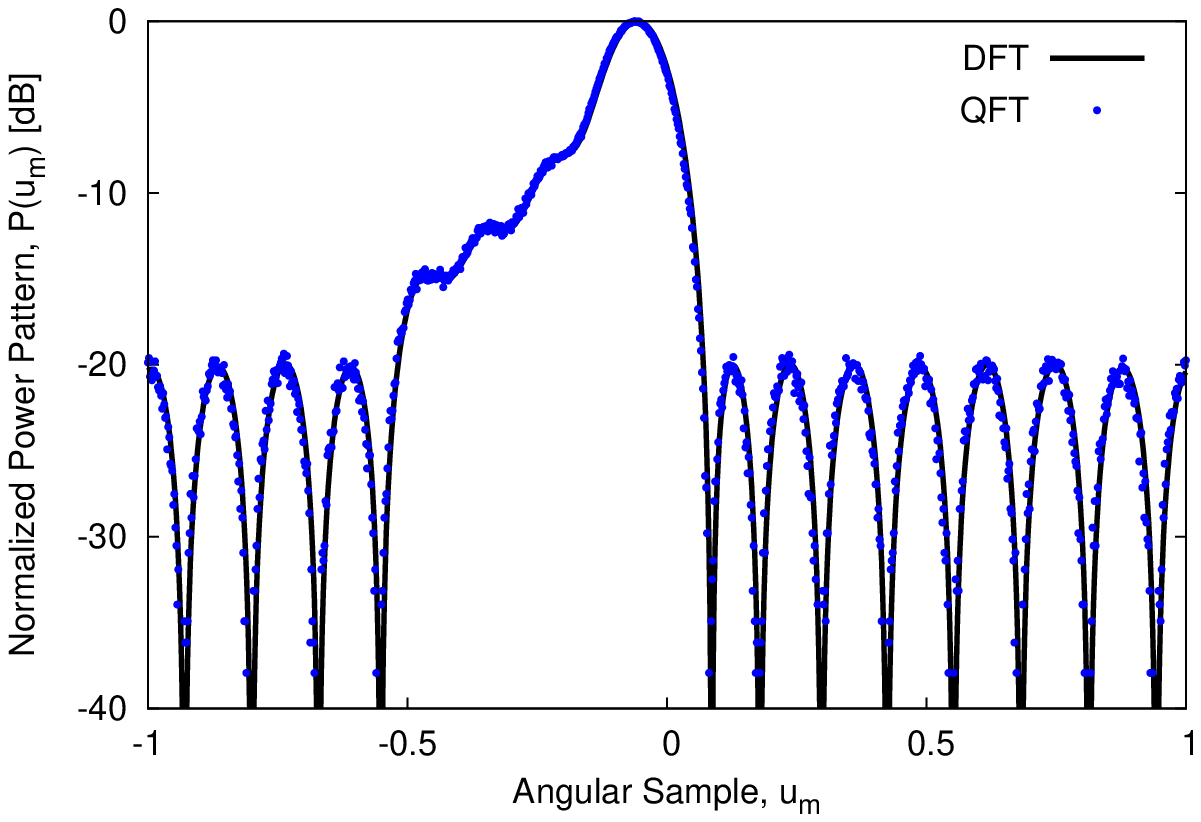}}\tabularnewline
\textcolor{black}{(}\textcolor{black}{\emph{e}}\textcolor{black}{)}&
\textcolor{black}{(}\textcolor{black}{\emph{f}}\textcolor{black}{)}\tabularnewline
\end{tabular}\end{center}

\begin{center}\textcolor{black}{~\vfill}\end{center}

\begin{center}\textbf{\textcolor{black}{Fig. 12 - L. Tosi et}} \textbf{\textcolor{black}{\emph{al.}}}\textbf{\textcolor{black}{,}}
\textcolor{black}{{}``Array Antenna Power Pattern Analysis ...''}\end{center}
\newpage

\begin{center}\textcolor{black}{~\vfill}\end{center}

\begin{center}\textcolor{black}{}\begin{tabular}{|c|c|c|c|c|}
\hline 
\textcolor{black}{\emph{Element Index, $n$}}&
\multicolumn{4}{c|}{\textcolor{black}{\emph{Normalized Excitation Module, $\left|\hat{w}_{n}\right|$}}}\tabularnewline
\hline
\hline 
&
\multicolumn{3}{c|}{\textcolor{black}{\emph{DC}}}&
\textcolor{black}{\emph{Taylor}}\tabularnewline
\hline
\hline 
&
\textcolor{black}{$SLL=-15\,[dB]$}&
\textcolor{black}{$SLL=-20\,[dB]$}&
\textcolor{black}{$SLL=-25\,[dB]$}&
\tabularnewline
\hline 
\textcolor{black}{$0$}&
\textcolor{black}{$0.4129$}&
\textcolor{black}{$0.2638$}&
\textcolor{black}{$0.1643$}&
\textcolor{black}{$0.2971$}\tabularnewline
\hline 
\textcolor{black}{$1$}&
\textcolor{black}{$0.1574$}&
\textcolor{black}{$0.1535$}&
\textcolor{black}{$0.1345$}&
\textcolor{black}{$0.2582$}\tabularnewline
\hline 
\textcolor{black}{$2$}&
\textcolor{black}{$0.1814$}&
\textcolor{black}{$0.1892$}&
\textcolor{black}{$0.1786$}&
\textcolor{black}{$0.2161$}\tabularnewline
\hline 
\textcolor{black}{$3$}&
\textcolor{black}{$0.2033$}&
\textcolor{black}{$0.2232$}&
\textcolor{black}{$0.2226$}&
\textcolor{black}{$0.2055$}\tabularnewline
\hline 
\textcolor{black}{$4$}&
\textcolor{black}{$0.2221$}&
\textcolor{black}{$0.2534$}&
\textcolor{black}{$0.2633$}&
\textcolor{black}{$0.2270$}\tabularnewline
\hline 
\textcolor{black}{$5$}&
\textcolor{black}{$0.2370$}&
\textcolor{black}{$0.2780$}&
\textcolor{black}{$0.2974$}&
\textcolor{black}{$0.2540$}\tabularnewline
\hline 
\textcolor{black}{$6$}&
\textcolor{black}{$0.2473$}&
\textcolor{black}{$0.2953$}&
\textcolor{black}{$0.3219$}&
\textcolor{black}{$0.2653$}\tabularnewline
\hline
\textcolor{black}{$7$}&
\textcolor{black}{$0.2526$}&
\textcolor{black}{$0.3043$}&
\textcolor{black}{$0.3347$}&
\textcolor{black}{$0.2640$}\tabularnewline
\hline
\end{tabular}\end{center}

\begin{center}\textcolor{black}{~\vfill}\end{center}

\begin{center}\textbf{\textcolor{black}{Tab. I - L. Tosi et}} \textbf{\textcolor{black}{\emph{al.}}}\textbf{\textcolor{black}{,}}
\textcolor{black}{{}``Array Antenna Power Pattern Analysis ...''}\end{center}
\end{document}